\documentclass[fleqn,usenatbib]{mnras}

\usepackage{newtxtext,newtxmath}

\usepackage[T1]{fontenc}
\usepackage{ae,aecompl}

\usepackage{graphicx}	
\usepackage{amsmath}	
\usepackage{mathrsfs}
\usepackage{blindtext}
\usepackage{rotating}
\usepackage{pdflscape}
\usepackage{adjustbox}
\usepackage{marvosym}
\usepackage{booktabs}

\newcommand{\LCDM}{$\Lambda$CDM\ }
\newcommand{\lya}{Ly$\alpha$\ }
\newcommand{\hi}{$\rm HI$\ }
\newcommand{\kpc}{\: \rm kpc}
\newcommand{\Mpc}{\: \rm Mpc}
\newcommand{\ckpc}{\: \rm ckpc}
\newcommand{\cMpc}{\: \rm cMpc}
\newcommand{\hMpc}{\: \mathrm{cMpc}/h}
\newcommand{\simba}{\textsc{Simba}}
\newcommand{\illustris}{\textsc{Illustris}}
\newcommand{\nyx}{\textsc{Nyx}}

\title[Impact of feedback on the CGM of quasars in \simba]{\textsc{Simba}: The average properties of the circumgalactic medium of $2 \leq z \leq 3$ quasars are determined primarily by stellar feedback}
\author[D. Sorini et al.]{
Daniele Sorini,$^{1}$\thanks{E-mail: sorini@roe.ac.uk}
Romeel Dav\'e$^{1, 2, 3}$ \&
Daniel Angl\'es-Alc\'azar$^{4, 5}$
\\
$^{1}$Institute for Astronomy, University of Edinburgh, Blackford Hill, Edinburgh, EH9 3HJ, United Kingdom\\
$^{2}$University of the Western Cape, Bellville, Cape Town 7535, South Africa\\
$^{3}$South African Astronomical Observatories, Observatory, Cape Town 7925, South Africa\\
$^{4}$Center for Computational Astrophysics, Flatiron Institute, 162 Fifth Avenue, New York, NY 10010, USA\\
$^{5}$Department of Physics, University of Connecticut, 196 Auditorium Road, U-3046, Storrs, CT 06269-3046, USA
}

\date{Accepted XXX. Received YYY; in original form ZZZ}

\pubyear{2020}

\begin{document}
\label{firstpage}
\pagerange{\pageref{firstpage}--\pageref{lastpage}}
\maketitle

\begin{abstract}
\noindent
We use the \simba\ cosmological hydrodynamic simulation suite to explore the impact of feedback on the circumgalactic medium (CGM) and intergalactic medium (IGM) around $2 \leq z \leq 3$ quasars. We identify quasars in \simba\ as the most rapidly-accreting black holes, and show that they are well-matched in bolometric luminosity and correlation strength to real quasars. We extract \lya absorption in spectra passing at different transverse distances ($10 \, \mathrm{kpc} \lesssim b \lesssim 10 \Mpc$) around those quasars, and compare to observations of the mean \lya absorption profile. The observations are well reproduced, except within $100 \, \kpc$ from the foreground quasar, where \simba\ overproduces absorption; this could potentially be mitigated by including ionisation from the quasar itself. By comparing runs with different feedback modules activated, we find that (mechanical) AGN feedback has little impact on the surrounding CGM even around these most highly luminous black holes, while stellar feedback has a significant impact. By further investigating thermodynamic and kinematic properties of CGM gas, we find that stellar feedback, and not AGN feedback, is the primary physical driver in determining the average properties of the CGM around $z\sim 2-3$ quasars. We also compare our results with previous works, and find that \simba\ predicts much more absorption within $100 \kpc$ than the \nyx\ and \illustris\ simulations, showing that the \lya absorption profile can be a powerful constraint on  simulations. Instruments such as VLT-MUSE and upcoming surveys (e.g., WEAVE and DESI) promise to further improve such constraints.

\end{abstract}




\section{Introduction}

Understanding the evolution of diffuse gas in the Universe is an essential prerequisite for a satisfactory theory of galaxy formation in a cosmological context. Indeed, about 90\% of baryons at $z\sim 2-3$ reside in a pervasive gaseous medium filling intergalactic space \citep[see, e.g.,][and references therein]{Rauch_1998}, known as intergalactic medium (IGM; see \citealt{Meiksin_review} and \citealt{McQuinn_2016} for reviews), which thus represents a gas reservoir for forming galaxies. Moreover, the gaseous environment at the interface of the IGM and galaxies, i.e. the circumgalactic medium (CGM), plays a pivotal role in the build up of galaxies, since crucial processes such as gas accretion and feedback-driven outflows are most prominent within the CGM \citep[see][and \citealt{Tumlinson_review} for recent reviews]{CGM_review}. It is then clear how the physics of gas encompasses an expansive range of scales, stretching from the filaments of the cosmic web down to sub-galactic regions.

Absorption lines in the spectra of background quasars (QSOs) represent an exquisite observational probe of the diffuse gas in the intervening IGM, and in the CGM of foreground galaxies at small transverse separations from the line of sight (LOS). For instance, an excess of neutral hydrogen (HI) absorption with respect to the IGM was observed in the CGM of foreground star-forming galaxies in the redshift range $2 \lesssim z \lesssim 3$ from the observations of 15 very luminous QSOs in the Keck Baryonic Structure Survey (KBSS) \citep[][and references therein]{Steidel_2010, Rakic_2012, Rudie_2012_CGM, Rudie_2013}. This result was subsequently confirmed by \cite{Turner_2014}, who also detected higher optical depth for metal lines close to galaxies. Later, systematic studies of \lya absorbers with high optical depth in the IGM at $2.6\lesssim z \lesssim 3.3$ revealed overdensities in the cosmic web on scales $\sim 10-20 \Mpc$, thus constraining structure formation models \citep{MAMMOTH_1, MAMMOTH_2}. More recently, \lya forest tomography techniques \citep{Pichon_2001, Caucci_2008, Gallerani_2011, Stark_2015_1, Stark_2015_2, Lee_2016, TARDIS} enabled the 3D reconstruction of the cosmic web thanks to various surveys (e.g., CLAMATO \citealt{CLAMATO_pilot, CLAMATO_DR1}, LATIS \citealt{LATIS}, and eBOSS \citealt{Ravoux_2020}), whereby \lya absorption in spectra of $z\sim2-3$ galaxies and quasars is utilised as a probe of diffuse gas in the intervening IGM, and around foreground star-forming galaxies and protoclusters (\citealt{Lee_colossus}; see also \citealt{Mukae_2019} and \citealt{Momose_2020_obs} for related studies).

QSOs are a particularly interesting class of objects to explore with absorption lines, given that their CGM are likely experiencing strong AGN (active galactic nucleus) feedback. The \lya absorption line was exploited to investigate the CGM around QSOs for the first time with the Quasars Probing Quasars (QPQ) project \citep[][and references therein]{QPQ10}, which consisted in the observation of a large sample of projected QSO pairs with small transverse separation ($< 1\Mpc$) at $z\sim 2-3$ (\citealt{Hennawi_2004, 
Hennawi_2006_binary_paper, Hennawi_2010, Hennawi_2006,QPQ2}; see also \citealt{Bowen_2006, Farina_2011, Farina_2013,Johnson_2013,Farina_2014,Johnson_2015,Johnson_2015_environment,Johnson_2016} for similar
works at lower redshifts). As part of this observational campaign, \cite{Prochaska_2013} observed an enhanced \lya absorption within $1 \Mpc$ from foreground QSOs \citep[see also][]{Prochaska_2013_QPQ5}, due to the presence of \hi and metals (\citealt{QPQ7,Lau_2016}; see also \citealt{QPQ9}), revealing a considerable reservoir of cool  ($T\sim 10^4 \, \rm K$) and metal-enriched gas \citep{Prochaska_2013_QPQ5}.

Using QSO spectra in the redshift range $2\lesssim z \lesssim 3$ from the Baryonic Oscillation Spectroscopic Survey (BOSS; \citealt{Ahn_2012}), \cite{Font-Ribera_2013} measured the \lya forest--QSO cross-correlation function. Such observations were later updated \citep{du_Mas_des_Bourboux_2017, Blomqvist_2019} with more recent data releases \citep{Alam_2015, Abolfathi_2018, Paris_2018}. The \lya--QSO cross-correlation is equivalent to the observable provided by \cite{Prochaska_2013} in the QPQ project, as shown for the first time by \cite{Sorini_2018}. Thus, BOSS/eBOSS enabled the extension of the QPQ \lya absorption profiles out to $80 \hMpc$ from the foreground QSOs, i.e. covering three decades in transverse distance. In an analogous manner, \cite{Perez-Rafols_2018} used BOSS/eBOSS quasar spectra  at $2\lesssim z \lesssim 3$ to also measure the cross-correlation between \lya forest and damped \lya absorbers (DLAs), superseding the previous observations by \cite{Font-Ribera_2012b}. These measurements can be converted into a \lya absorption profile too, and as such they constitute an extension to larger scales of \cite{Rubin_2015} observations of close QSO pairs, whereby one line served for the identification of foreground DLAs, and the other one as a probe of \lya and metal line absorption at transverse distances $< 200 \kpc$.

All aforementioned absorption-line observations provide an effective way to trace the composition of IGM gas in the Universe. In particular, the abundance and ionisation state of HI within the IGM is set by the balance of  photoionisation due to UV photons emitted by galaxies and QSOs, and of HI recombination, which is determined by the local density and temperature of the gas \citep{Meiksin_review,McQuinn_2016}. The physics is more complex within the CGM of galaxies and QSOs, where higher densities and temperatures make HI self-shielding non-negligible, and enable further ionisation processes, such as collisional ionisation. Moreover, galactic winds and outflows driven by the central AGN impact the properties of the gas in the CGM, which thus represents the link between galaxies and the large-scale structure of the IGM. As such, to achieve a consistent physical description of diffuse gas in the Universe and particularly in the CGM, it is imperative to fully model galaxy formation processes embedded in a cosmological context.

Given the non-linear and multi-scale nature of the evolution of IGM/CGM and galaxies, it is essential to rely on cosmological hydrodynamic numerical simulations. While such simulations represent the best effort to capture all relevant physical processes, they are often time expensive and memory intensive. In fact, due to numerical constraints, designing a cosmological hydrodynamic simulation always requires a trade-off between volume and resolution. For this reason, fundamental physical processes on galactic scales such as feedback from winds driven by supernovae or AGN jets are often implemented in the form of simulation-specific sub-grid prescriptions \citep[see][for a review]{Somerville_review}. The reliability of any given feedback prescription is generally validated a posteriori by verifying that the simulation successfully reproduces different sets of observations, for instance the stellar mass function \citep{Baldry_2008, Baldry_2012, Bernardi_2013, D_Souza_2015}, the gas fraction within haloes \citep[e.g.,][]{Giodini_2009, Lovisari_2015}, the star formation efficiency \citep{Guo_2011, Behroozi_2013, Moster_2013}, or the evolution of the star formation rate density \citep{Behroozi_2013, Oesch_2015}.

A complementary set of constraints on feedback prescriptions can be obtained by comparing the predictions of cosmological simulations with the aforementioned observations of absorption lines in the CGM and IGM, particularly considering the ever increasing precision of such measurements thanks to recent and upcoming surveys (e.g., BOSS, WEAVE \citealt{WEAVE}, DESI \citealt{DESI}). Obviously, because feedback prescriptions always affect the stellar mass of galaxies, they also impact the correlation between stellar mass and absorption properties in the CGM, even when they would not directly affect the absorption properties themselves. It is however useful to compare the effect of different feedback mechanisms on the absorption properties of the CGM with respect to a fixed set of haloes, as DM haloes are generally only weakly affected by feedback. Within this setting, investigating the effect of stellar and AGN feedback on the properties of the CGM and IGM has a dual purpose: on one side, gaining further physical insight on their evolution, and on the other hand refining feedback prescriptions in the next generation of simulations from the constraints provided by the observations of these gaseous media.

The majority of past numerical studies of the CGM were mainly concerned with reproducing the covering factor of optically thick absorbers around galaxies and QSOs in the redshift range $z\sim 2-3$. While recent simulations \citep{Ceverino_2012, Dekel_2013, Shen_2013,Meiksin_2015, Suresh_2015,Meiksin_2017, Suresh_2019} were able to broadly reproduce \cite{Rudie_2012_CGM} measurements of this quantity around galaxies, the high covering factor observed around QSOs by \cite{Prochaska_2013} proved to be harder to reproduce \citep{Fumagalli_2014, Faucher-Giguere_2015}. Later, \cite{Faucher-Giguere_2016} was able to recover such measurements with the FIRE zoom-in simulations \citep{Hopkins_2014}, which included only stellar feedback, arguing that high resolution was a crucial element to obtain this result. However, \cite{Rahmati_2015} succeeded in reproducing these data with the EAGLE \citep{Crain_2015, Schaye_2015} suite of cosmological hydrodynamic simulations, implementing both stellar and AGN feedback, at much lower resolution. Considering this debate about resolution and the different feedback prescriptions involved, reproducing absorption-line observations around QSOs still remains an important issue.

Another body of work focussed on the column density distribution function (CDDF) of \hi absorbers at high redshift ($z\sim 2-3$).  Measurements of this quantity \citep{Kim_2002, Peroux_2005, Zwaan_2005, O_Meara_2007, Noterdaeme_2009,  Prochaska_2009b, Prochaska_2010, Noterdaeme_2012, Kim_2013, Rudie_2013} were mostly successfully reproduced in several different simulations \citep{Altay_2011, Altay_2012, Rahmati_2013, Rahmati_2013b, Rahmati_2015, Bird_2013, Bird_DLA, Fumagalli_2011,McQuinn_2011_CDDF}. Indeed, matching this CDDF was a natural outcome even in the earliest generation of cosmological simulations that included no feedback at all \citep[e.g.][]{Dave_1997}.  At low redshift ($z<0.2$), observations of the distribution of \hi column density around galaxies \citep{Prochaska_2011, Tumlinson_2013, Prochaska_2017} were reproduced by \cite{Gutcke_2017} with the NIHAO \citep{Wang_2015} suite of zoom-in simulations \citep[see also][]{Stinson_2012}, by \cite{Hafer_2017} with the FIRE simulations, and by \cite{van_de_Voort_2019} within the Auriga project \citep{Auriga}. On the other hand, simulations of idealised isolated galaxies \citep{Butsky_2018} struggled reproducing analogous observations by \cite{Werk_2013} over the full range of transverse distance ($0-200 \kpc$), highlighting the importance of simulating the evolution of galaxies within full cosmological simulations.

Directly related to the \hi content of CGM and IGM surrounding galaxies, the average \lya absorption profile is another very well-studied statistic. Large-volume hydrodynamic simulations with various feedback implementations have been employed in several works \citep[][see also \citealt{Chung_2019} for a related study with zoom-in simulations]{Kollmeier_2003, Kollmeier_2006, Rakic_2012, Rakic_2013, Meiksin_2014, Meiksin_2015, Meiksin_2017, Turner_2017} aiming at reproducing measurements of the \lya flux decrement around LBGs \citep{Adelberger_2003, Adelberger_2005, 
Steidel_2010,Crighton_2011, Rakic_2012, Turner_2014} and/or QSOs \citep{Prochaska_2013}. Except for data points within the virial radius of foreground objects, these observations were generally matched by the simulations. Using the Sherwood \citep{Bolton_2017} suite of simulations, \cite{Meiksin_2017} suggested that the discrepancy with observations close to the foreground objects could be mitigated by stronger stellar feedback, while \cite{Turner_2017} found that the \lya optical depth given by the EAGLE simulations depends only weakly on the stellar feedback model.

More recently, \cite{Sorini_2018} expanded this line of research by comparing the predictions of different cosmological simulations with several observations of \lya transmission at redshift $z\sim 2-3$ around QSOs \citep{Font-Ribera_2013, Prochaska_2013}, LBGs \citep{Adelberger_2003, Adelberger_2005, Crighton_2011, Turner_2014}, and DLAs \citep{Font-Ribera_2012b,Rubin_2015}, covering three decades of distance $(10 \, \mathrm{kpc} - 10 \Mpc)$ around such objects \citep[see also][]{Sorini_PhD}. Specifically, they employed the publicly available fiducial run of the \illustris\ cosmological simulation \citep{Vogelsberger_2014N, Vogelsberger_2014, Genel_2014, Illustris_public, Sijacki_2015}, and a large-volume and high-resolution run of the \nyx\ hydrodynamic code \citep{Almgren_2013, Lukic_2015}. The former is equipped with both stellar and AGN feedback, while the latter has no feedback implementation, and acts as a convenient reference run. \cite{Sorini_2018} further considered two variants of the \nyx\ run, whereby the effects of feedback were mimicked in post-processing with a semi-analytic model that allowed altering the temperature of the CGM of the haloes selected in the simulation to reproduce the observations of interest. The main result was that, while all simulations converged to the same predictions of the \lya transmission profiles at large transverse distance from foreground objects ($> 2 \Mpc$), and successfully reproduced the observations in this regime, there were discrepancies among simulations, and between simulations and data, on smaller scales. 

In this work, we revisit the \cite{Sorini_2018} study by addressing its main limitation: the lack of a unique suite of simulations, run with exactly the same code, and differing solely by the implementation of stellar and AGN feedback. We do this by using six different runs of the \simba\ suite of cosmological hydrodynamic simulations \citep{Simba}, by means of which we explore the effect of stellar feedback, and of various AGN feedback models on the \lya absorption profile around QSOs at $z \sim 2-3$, and on the thermodynamic properties of the surrounding gaseous environment. We also compare the predictions of \simba\ with the results previously obtained by \cite{Sorini_2018}. We find that all \simba\ runs broadly agree with \nyx\ and \illustris\ on large scales ($\gtrsim 2 \Mpc$), but it predicts significantly higher \lya absorption within $100 \kpc$ from QSOs. This confirms the constraining power of the \lya absorption profile: the increase of precision in data due to ongoing and future surveys (e.g. WEAVE, DESI) will soon enable to discriminate among the predictions of the different simulations. Our results from \simba\ show that stellar feedback is the dominant physical driver in determining the average physical properties of $z\sim 2-3$ QSOs, and consequently their \lya absorption properties, while the effect of AGN feedback is marginal. Unlike \cite{Sorini_2018}, in this paper we focus exclusively on the gaseous environment of QSOs, leaving the investigation of the \lya transmission around LBGs and DLAs for future work.

This manuscript is organised as follows. In \S~\ref{sec:simulations} we describe the main features of the simulations adopted in this work. In \S~\ref{sec:modelling} we explain how we model \lya absorption and how we reproduce the observations considered in this work from the simulations. In \S~\ref{sec:results} we present our results, and in \S~\ref{sec:discussion} we discuss the implications for the physics of the gas surrounding $z \sim 2-3$ QSOs. Finally, in \S~\ref{sec:conclusions} we state our conclusions and outline the perspectives of this work. Throughout this paper, distances are expressed in physical units (e.g., $\kpc$, $\Mpc$, etc.) unless otherwise indicated. When referring to co-moving units, we prefix the symbol of the unit of measure with a ``c'' (e.g., $\ckpc$, $\cMpc$, etc.).

\section{Simulations}
\label{sec:simulations}

In this work, we adopt several runs of the \simba\ simulation for our computations. We summarise its main features in \S~\ref{sec:simba}, where we also provide specific details of the runs considered. Since we will compare our results from \simba\ with those obtained by \cite{Sorini_2018} with \illustris\ and \nyx, we briefly describe these simulations in \S~\ref{sec:illustris} and \S~\ref{sec:nyx}, respectively.

\subsection{Simba}
\label{sec:simba}

\simba\ \citep{Simba} is a hydrodynamic cosmological simulation built upon its predecessor \textsc{Mufasa} \citep{Mufasa}. Dark matter (DM) is treated with a Lagrangian approach, while gas is evolved following the meshless finite mass (MFM) implementation of the \texttt{Gizmo} hydrodynamic code \citep{Gizmo}, which enables an accurate description of shocks and shear flows, without introducing any artificial viscosity \citep{Gizmo}. This feature thus allows us to faithfully follow flows with high Mach number and shocks, as it is the case for outflows and jets.

Radiative cooling and photoionisation heating are implemented through the \texttt{Grackle-3.1} library \citep{Smith_2017}, which accounts for metal cooling and non-equilibrium evolution of primordial elements. The UV ionising background (UVB) follows the \cite{HM12} model, modified to account for self-shielding self-consistently throughout the simulation run, according to the \cite{Rahmati_2013} prescription (A. Emerick, priv. comm.).
 This improves the accuracy of the thermodynamic properties of circumgalactic gas. The neutral hydrogen content of gas particles is computed self-consistently on the fly, and not by applying self-shielding in post-processing \citep{Dave_2017b}. Star formation is modelled following a Kennicutt-Schmidt law \citep{Kennicutt_1998}, scaled by the $\rm H_2$ fraction, determined from the local column density and metallicity of the gas particle according to the variant of \cite{Krumholz_2011} sub-grid model discussed in \cite{Mufasa}. The chemical enrichment model allows tracking eleven different elements  (H, He, C, N, O, Ne, Mg, Si, S, Ca, Fe) from Type Ia and II supernovae (SNe), and Asymptotic Giant Branch (AGB) stars. For SNIa and SNII, this is done by adopting the yield tables given by \cite{Iwamoto_1999} and \cite{Nomoto_2006}, respectively, while for AGB stars by following the chemical enrichment model by \cite{Oppenheimer_2006}. Star formation can occur only above the hydrogen density threshold $n_{\rm H} \geq 0.13 \, \rm cm^{-3}$. Gas above such threshold is considered ``interstellar medium'' (ISM), and is subject to an artificial pressurisation scheme in order to resolve the Jeans mass \citep[see][]{Mufasa}.

Star formation-driven galactic winds are modelled in a two-phase fashion, where the temperature of 30\% of the wind particles ejected is set by the supernova energy minus the kinetic energy of the wind. The mass loading factor scales following the outflow rates found by \cite{Angles-Alcazar_2017b} within the FIRE zoom-in simulations. Winds are metal-loaded, and their metallicity is set by the Type II SNe yields and the mass loading factor. The velocity scaling of winds follows that found by \cite{Muratov_2015} from the FIRE simulations. 

\simba\ includes BH particles, which accrete following a dual model. The hot-accretion mode follows the Bondi accretion from the hot gas component. The cold-accretion mode is described with a torque-limited accretion model, driven by disk gravitational instabilities arising from galactic scales down to the accretion disk around the central BH (\citealt{Hopkins_2011}; see also \citealt{Angles-Alcazar_2013, Angles-Alcazar_2015, Angles-Alcazar_2017a}). 

\subsubsection{AGN feedback}
\label{sec:feedback}

AGN feedback is implemented in \simba\ through three different modes, which we summarise in this section.
\begin{itemize}
	\item \textit{AGN winds}: BHs with high accretion rate ($>0.2$ times the Eddington accretion rate) eject purely bipolar outflows, the velocity of which scales logarithmically with the BH mass. The winds are kinetically coupled to the surrounding gas, without changing its temperature, which is set by the ISM pressurisation model. This is consistent with observations of ionised gas outflows, which suggest electron temperatures of order $10^4 \, \rm K$ \citep[e.g.][]{Perna_2017b}.
	\item \textit{Jets}: When the BH accretion rate drops below $0.2$ times the Eddington accretion rate and the mass of the BH exceeds $10^{7.5}  \, \rm M_{\odot}$\footnote{This is a conservative mass cut motivated by observations of jets arising only in galaxies with velocity dispersions consistent with a BH mass of $\gtrsim 10^8 \, M_{\odot}$ \citep{Barisic_2017}.}, AGN feedback begins a transition to jet mode. Jets are still implemented in the form of outflowing perfectly bipolar winds kinetically coupled to the gas surrounding the BH. In addition to the velocity determined by the \textit{AGN winds} feedback mode, jets receive a velocity increment proportional to the logarithm of the inverse of the accretion rate in units of the Eddington accretion rate. Such increment is capped at $7000 \, \rm km/s$.  Full jet mode is achieved when the BH accretion rate drops below 0.02 of Eddington.
	\item \textit{X-ray heating}: This is activated if a BH satisfies the criteria for the \textit{Jets} feedback mode, and the gas fraction of the host galaxy is below $0.2$. Only gas within the BH kernel is subject to X-ray heating, which is proportional to the inverse square of the distance of the gas element with respect to the BH. \footnote{This includes the Plummer softening based on the smoothing scale of gas, to prevent excessively large deposition of energy in gas in the immediate vicinity of the BH.} Non-ISM gas is heated by directly increasing its temperature according to the heating flux at the position of the gas particle. For ISM gas, half of the X-ray energy is applied kinetically as a radial outwards kick, and the other half is added as heat. This prescription prevents quick cooling in the low-resolution ISM, which would occur by the ISM pressurisation model of \simba\ \citep{Mufasa}. 
\end{itemize}

\subsubsection{Runs}

\begin{table*}
\begin{center}
\begin{tabular}{lccccccc}
\hline
Simulation & Box size ($\hMpc$) & Nr. of particles & $\Gamma_{\rm UVB}/\Gamma^{\rm HM12}_{\rm UVB}$ & 
Stellar Feedback & AGN winds & Jets & X-ray heating \\
\hline
Simba $100 \hMpc$ & 100  & $2\times 1024^3$ & 2.0240 & \checkmark  & \checkmark  & \checkmark & \checkmark  \\
Simba $50 \hMpc$ & 50  & $2\times 512^3$ & 1.9744 & \checkmark  & \checkmark  & \checkmark & \checkmark  \\
Simba $25 \hMpc$ & 25  & $2\times 512^3$ & 1.9496 & \checkmark  & \checkmark  & \checkmark & \checkmark  \\
SFB + AGN Winds + Jets & 50  & $2\times 512^3$ & 1.9250 & \checkmark  & \checkmark  & \checkmark  & \\
SFB + AGN Winds & 50  & $2\times 512^3$ & 1.9994 & \checkmark  & \checkmark & & \\
Stellar Feedback & 50  & $2\times 512^3$ & 1.9998 & \checkmark & & & \\
No Feedback & 50  & $2\times 512^3$ & 1.9378 & & & & \\
\hline
\end{tabular}
\caption{\simba\ runs used in this work. The fourth column from the left shows the factor applied to the \protect\cite{HM12} UVB in order to match the mean flux at $z=2.4$ observed by \protect\cite{Becker_2013_mean} (see \S~\ref{sec:skewers} for details).
}
\label{tab:runs}
\end{center}
\end{table*}

In this work, we use six runs of the \simba\ suite of hydrodynamic simulations. Our fiducial run is a $100 \hMpc$ box with $1024^3$ DM particles and as many gas particles, with a mass resolution of $9.6 \times 10^7 \, M_{\odot}$ and $1.82 \times 10^7 \, M_{\odot}$, respectively. All physical prescriptions described earlier in this section are implemented in this run. The simulation is built upon a \LCDM cosmological model consistent with \cite{Planck_2016} cosmological parameters ($\Omega_{\mathrm{m}}=0.3$, $\Omega_{\Lambda}=1-\Omega_{\mathrm{m}}=0.7$,
$\Omega_{\mathrm{b}}=0.048$, $h=0.68$, $\sigma_8=0.82$, $n_s=0.97$, with the usual definitions of the parameters).

To test the effect of stellar feedback and of the different AGN feedback modes on the properties of the IGM and CGM surrounding $z\sim 2-3$ QSOs, we also consider five runs with a $50 \hMpc$ box and $2\times 512^3$ DM and gas particles, with the same mass resolution as the the fiducial simulation. One run has no feedback prescription at all, in another one we include stellar feedback, but none of the AGN feedback prescriptions described in \S~\ref{sec:feedback}, while in the remaining three runs we activate only the first, first two, and all three AGN modes, respectively. In all plots in this manuscript, we will refer to the various runs with the labels defined in Table~\ref{tab:runs}. In the main text, we will also refer to the runs with stellar feedback and all AGN feedback modes as ``full \simba'' runs, always specifying their box size to avoid any ambiguity. All $50 \hMpc$ \simba\ runs differ only by the number of AGN feedback modes implemented; they are otherwise identical, and start with the same initial conditions. In particular, all runs include accreting BHs, and the star formation prescriptions, including metal enrichment, are the same across all runs. This implies that observables such as the mass-metallicity relation are not reproduced in all runs (e.g. the no-feedback run).

The \simba\ $50 \hMpc$ run relies on the same physics implemented in its $100 \hMpc$ counterpart. We also used a smaller variant of the full \simba\ run ($25 \hMpc$, $2\times 512^3$ particles) exclusively for convergence tests (see appendix \S~\ref{sec:convergence}). We could not explore the various AGN feedback prescriptions in a suite of $100 \hMpc$ \simba\ simulations with $2\times 1024^3$ particles, as we did for the $50 \hMpc$ runs, because of the computational resources available. 

During each run, haloes are identified on the fly via a 3D friends-of-friends algorithm embedded in \texttt{Gizmo}, taken from the one written by V. Springel in \texttt{Gadget-3}, using 0.2 times the mean interparticle separation as linking length. Galaxies and haloes are cross-matched in post-processing with the \texttt{yt}-based package \textsc{Caesar} \footnote{\url{https://caesar.readthedocs.io/en/latest/}}, which generates a catalogue with several key pre-computed properties. Our results are obtained from the \textsc{Caesar} catalogues corresponding to the snapshots of interest.  We will describe the generation of \lya\ absorption spectra in \S\ref{sec:skewers}.

\subsection{Illustris}
\label{sec:illustris}

\illustris\ \citep{Vogelsberger_2014, Vogelsberger_2014N, Genel_2014, Sijacki_2015} is a cosmological hydrodynamic simulation run with the \texttt{Arepo} code  \citep{Springel_2010}. Dark matter is described as a set of Lagrangian particles, and baryons are represented by an ideal gas on a moving mesh derived from a Voronoi tessellation of the simulation box. Gravitational forces are calculated following a Tree-PM scheme \citep{Xu_1995}, with long-range  and short-range forces computed through a particle-mesh method and a hierarchical algorithm \citep{Barnes_1986}, respectively. Gas evolution is followed via the viscosity-free Euler equations.

The simulation accounts for several astrophysical processes, such as  primordial and metal-line cooling, gas recycling and chemical enrichment. \illustris\ also includes a sub-resolution model of the interstellar medium, stochastic star formation above a density threshold of $0.13 \: \rm cm^{-3}$, supermassive black hole seeding, accretion  and merging \citep[see][for details]{Vogelsberger_2013}. Feedback from AGN is implemented through a dual modelling \citep{Sijacki_2007}, based on the BH accretion rate. For high accretion rates, a ``quasar-mode'' AGN feedback is activated, whereby the energy radiated by the BH is thermally coupled to the surrounding gas. For slowly accreting BHs, hot gas bubbles are injected in the halo atmosphere via a mechanical ``radio-mode'' AGN feedback. The free parameters underlying feedback prescriptions were tuned to reproduce the overall observed star formation efficiency \citep{Guo_2011,Moster_2013,Behroozi_2013} in a set of smaller-scale simulations  \citep{Vogelsberger_2013}.

Heating and photoionisation are computed from the UVB model by \cite{Faucher-Giguere_2009}. Self-shielding in dense regions is included on the fly following \citealt{Rahmati_2013}.  Ionisation from neighbouring AGN are included in the computation of cooling and heating of gas cells. 

The initial redshift of the simulation is $z_{\rm ini}=127$ \cite[see][for details]{Vogelsberger_2014N}. The \LCDM cosmological model is consistent with the parameters obtained in the 9-year data release of WMAP \citep{Hinshaw_2013}: $\Omega_{\mathrm{m}}=0.2726$, $\Omega_{\Lambda}=1-\Omega_{\mathrm{m}}=0.7274$,
$\Omega_{\mathrm{b}}=0.0456$, $h=0.704$, $\sigma_8=0.809$, $n_s=0.963$. In this work, we will consider the results obtained by \cite{Sorini_2018} with the snapshot at $z=2.44$ of the ``Illustris-1'' run, i.e. the one with the highest resolution available. The simulation size is $75 \hMpc$ per side; there are $1820^3$ DM particles, and as many gas Voronoi cells. As such, the mean inter-particle separation is $58.5 \ckpc$. 
The mass resolution is $6.3\times10^6\: M_{\odot}$ and $1.3\times10^6\: M_{\odot}$ for DM and gas, respectively.

\subsection{Nyx}
\label{sec:nyx}

To compare our findings with \simba\ to the predictions of a feedback-free model operating on a totally different code, we will consider the results obtained by \cite{Sorini_2018} with the \nyx\ \citep{Almgren_2013, Lukic_2015}. \nyx\ treats DM as self-gravitating Lagrangian particles, and baryonic matter as an inviscid fluid that obeys an equation of state resembling that of an ideal gas. Eulerian equations of gas dynamics are solved on a regular Cartesian grid. The Riemann problem is solved iteratively, following a second-order-accurate piece-wise parabolic method \citep{Colella_1985}, which ensures accurate description of shock waves.

Gas is assumed to have a primordial composition, with hydrogen and helium abundances $X_{\rm p}=0.76$ and $Y_{\rm p}=0.24$, respectively. Inverse-Compton cooling off the microwave background and thermal energy loss due to atomic collisional processes are included. The values of the recombination, collisional ionization, dielectric recombination rates, and cooling rates in the \nyx\ run used by \cite{Sorini_2018} can be found in \cite{Lukic_2015}. The UVB model follows \cite{HM12}.

Star formation is not implemented in \nyx. As a consequence, the central regions of haloes exhibit artificially high densities and low temperatures. To circumvent this issue, \cite{Sorini_2018} imposed a ceiling of $\delta = 1000$ to the gas overdensity when computing \lya mock absorption spectra (see the original paper for further details). Neither stellar nor AGN feedback are included in \nyx.

In this work, we report the results from the $z=2.4$ snapshot of the \nyx\ run analysed by \cite{Sorini_2018}. The simulation volume is $(100 \hMpc)^3$, with a grid of $4096^3$ gas cells and as many DM particles. The resolution of $35.6 \, \ckpc$ for baryons guarantees a precision within $5\%$ in the 1D power spectrum, and at percent level in the probability density function (PDF), of the \lya forest flux \citep{Lukic_2015}. The simulation is initialized at redshift $z_{\rm ini} = 200$, ensuring that non-linear evolution is not compromised \citep[for a detailed discussion see, e.g.,][]{Onorbe_2014}. Cosmology follows a $\Lambda$CDM model with parameters consistent with \citep{Planck_2016}: $\Omega_{\mathrm{m}}=0.3$, $\Omega_{\Lambda}=1-\Omega_{\mathrm{m}}=0.7$, $\Omega_{\mathrm{b}}=0.047$, $h=0.685$, $\sigma_8=0.8$, $n_s=0.965$. The adaptive mesh refinement feature is not active in the run considered. \cite{Sorini_2018} incorporated self-shielding in the computation of \lya optical depth, following \cite{Rahmati_2013} formula. We refer the interested reader to the original papers for further details.

\section{Modelling}
\label{sec:modelling}

We want to investigate the mean \lya absorption profile around QSOs in \simba, and compare it with the observations by \cite{Prochaska_2013} and \cite{Font-Ribera_2013}. To do this, we first need to select a sample of objects acting as QSOs from the simulation, and then generate \hi absorption spectra at different transverse distances from such objects. We describe these two aspects of our modelling in \S~\ref{sec:sel_QSO} and \S~\ref{sec:skewers}, respectively.

\subsection{Selection of QSOs in \simba}
\label{sec:sel_QSO}

\begin{figure*}
	\includegraphics[width=\textwidth]{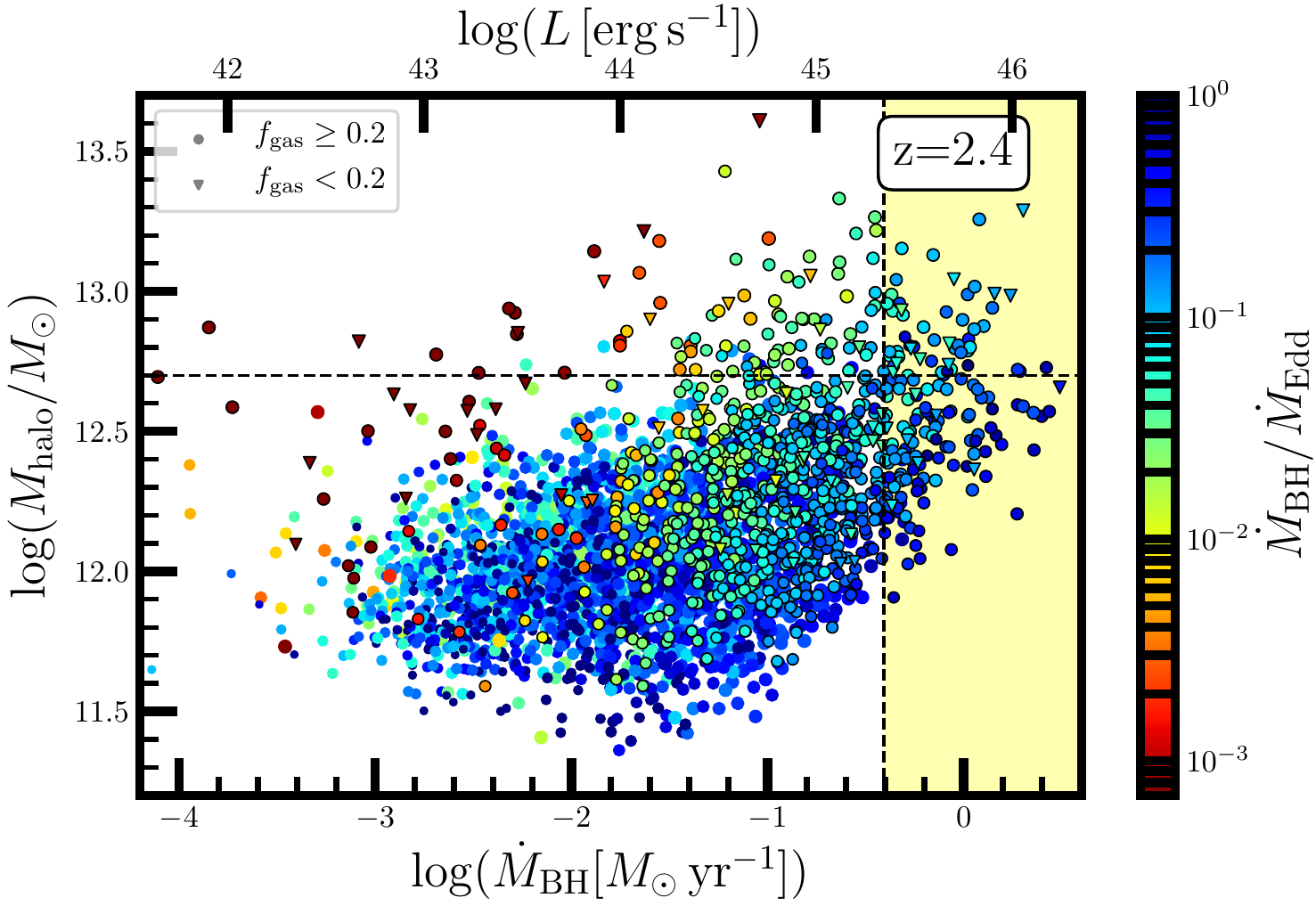}
    \caption{Host halo mass--central BH accretion rate relationship for central galaxies in the fiducial \simba\ $100 \hMpc$ run at $z=2.4$. The luminosities corresponding to the BH accretion rates are reported in the upper $x$-axis, and are deduced assuming the canonical value of 0.1 for the radiative efficiency. Central galaxies are plotted as circles if their gas mass fraction is at least 0.2, otherwise as reversed triangles. The color coding represents the BH accretion rate in units of the Eddington accretion rate. The size of the markers is proportional to the BH mass, and markers with a black edge correspond to BHs with mass exceeding $10^{7.5} \, \rm M_{\odot}$. In this way, the colour, shape, and size of any given marker enable us telling whether the corresponding BH exhibits AGN feedback activity, and if so, in which modes (see \S~\ref{sec:feedback} for details). The horizontal and vertical dashed lines represent, respectively, the host halo mass and luminosity cuts that need to be applied to the haloes within the simulation in order to obtain the best match to the QSO correlation function measured by \protect\cite{White_2012}, as explained in \S~\ref{sec:haloes}. The highlighted area at the right of the vertical dashed line identifies the QSO sample selected.}
    \label{fig:sel_haloes}
\end{figure*}

The definition of a sample of QSOs within a simulation is often accomplished by imposing specific selection criteria on their host haloes. 
For instance, one possibility is considering haloes within a certain mass range to be QSO hosts \citep[see, e.g.,][]{Meiksin_2014, Faucher-Giguere_2015, Rahmati_2015, Faucher-Giguere_2016, Meiksin_2015, Meiksin_2017}. Although this is a sensible choice, a mass-based selection criterion can become problematic when comparing the results of different simulations, which may not adopt the same halo-finding mechanism. More importantly, massive haloes in simulations are not a priori guaranteed to match any observed statistic of QSOs. For these reasons, \cite{Sorini_2018} calibrated the halo mass floor of QSO hosts such that the correlation function of the resulting sample of haloes matched the observations of the QSO correlation function by \cite{White_2012}. This method provides a mass-based selection criterion which is physically well motivated, although it effectively relies on the somewhat unrealistic assumption that the halo occupation distribution (HOD) of QSOs is a step function (but see  also \citealt{Rodriguez-Torres_2017}, and the discussion in \citealt{Sorini_2018}). 

In this work, we adopt an even more realistic selection criterion. As a starting point, we follow \cite{Sorini_2018} and determine the halo mass floor $M_{\rm min}$ that best fits \cite{White_2012} observations of QSO clustering. Because \simba\ incorporates BH accretion, we then consider all central galaxies\footnote{The central galaxy of a halo is defined as the most massive galaxy within that halo. This does not necessarily mean that the position of the central galaxy coincides with the centre of the host halo (see the discussion in \S~\ref{sec:position} on the implications for this work). We considered only central galaxies to be suitable QSO candidates; including satellite galaxies has negligible impact on our results (see \S~\ref{sec:central}).} in the simulation that are endowed with a central BH. The $N$ such galaxies containing the $N$ fastest accreting BHs, where $N$ is the number of haloes with mass $\geq M_{\rm min}$, are defined to be QSO hosts. The QSOs are assumed to be located exactly at the centre of their host galaxy. 

Our selection criterion has the advantage of being based both on halo mass and BH accretion rate. As such, our technique has a stronger physical motivation, as real QSOs are characterised by a high BH accretion rate and strong clustering. Thus, unlike in \cite{Sorini_2018}, our method provides a good match to the observed QSO clustering properties \citep{White_2012} without assuming a purely mass-based HOD, which can be simplistic \citep{Beltz-Mohrmann_2020, Hadzhiyska_2020}. At the same time, the number of haloes that we select is the same that we would have selected if we had simply considered all haloes with mass above $M_{\rm min}$. Therefore, our results can be easily compared with other works where QSO hosts are selected solely according to their halo mass. We stress that in some simulations (e.g., \nyx) there are no BH particles, therefore imposing a mass cut for the selection of QSO hosts is probably the only viable choice.

Figure \ref{fig:sel_haloes} shows the results of our selection criterion when applied to the \simba\ $100 \hMpc$ run at $z=2.4$. Every point in the plot represents a central galaxy. The size of the points is proportional to the mass of the central BH, and points with a black edge correspond to BHs with mass above the $10^{7.5} \, \rm M_{\odot}$ threshold needed to activate AGN jets (see \S~\ref{sec:feedback}). Galaxies with gas fraction larger than 0.2 are plotted as circles, and as reversed triangles otherwise. Points are colour coded according to the accretion rate in units of the Eddington accretion limit. Therefore, the marker style and colour of each point can immediately tell how many AGN feedback modes are active in the corresponding galaxy (cf. \S~\ref{sec:feedback}). The $y$-axis shows the mass of the host halo, and the lower $x$-axis the accretion rate of the central BH. The upper $x$-axis displays the corresponding luminosity, calculated as
\begin{equation}
\label{eq:L}
    L = \varepsilon \dot{M}_{\rm BH} c^2 \, ,
\end{equation}
where $\dot{M}_{\rm BH}$ is the BH accretion rate, $c$ the speed of light, and $\varepsilon$ the radiative efficiency. We assumed the canonical value $\varepsilon=0.1$ \citep[see, e.g.,][]{Trakhtenbrot_2017}. The horizontal and vertical dashed lines show the halo mass and accretion rate thresholds obtained with our selection technique ($10^{12.7}\, \rm M_{\odot}$ and $0.4\, \rm M_{\odot}\, yr^{-1}$, respectively). Thus, all points in the highlighted area on the right side of the vertical dashed line are considered QSO hosts in this work. These are galaxies containing AGN with luminosities above $10^{45.3}\, \rm erg \, s^{-1}$. Such luminosity range is consistent with typical QSO luminosities \citep[][and references therein]{Shen_2020}. Hence, this represents a further validation of our selection method. 

\begin{table}
\begin{center}
\begin{tabular}{lcccc}
\hline
Simulation & \textbf{All QSOs} & \multicolumn{3}{c}{QSOs exhibiting}\\
 &  & AGN winds & Jets & Jets + X-Ray \\
\hline
Simba $100 \hMpc$ & \textbf{176} & 80 & 78 & 18 \\
Simba $50 \hMpc$ & \textbf{25} & 9 & 15 & 1 \\
SFB+AGN Winds+Jets &  \textbf{24} & 15 & 9 & 0 \\
SFB+AGN Winds & \textbf{23} & 23 & 0 & 0 \\
Stellar Feedback & \textbf{25} & 0 & 0 & 0 \\
No Feedback & \textbf{27} & 0 & 0 & 0 \\
\hline
\end{tabular}
\caption{AGN feedback modes active in the QSOs selected from the $z=2.4$ snapshots of the \simba\ runs considered in this work. QSOs are considered to be exhibiting the jet mode as soon as the BH accretion rate drops below the threshold of 0.2 Eddington, and not when jets reach their full speed (see \S~\ref{sec:feedback}). The QSOs are selected by applying a halo-mass-calibrated luminosity cut, as explained in \S~\ref{sec:sel_QSO} and in the appendix \S~\ref{sec:haloes}. We report the values of the luminosity cuts that we applied in the various runs in Table~\ref{tab:selection} (third column from the left).}
\label{tab:QSOs}
\end{center}
\end{table}

The halo mass thresholds used to set the luminosity cuts that we obtain in the  five $50 \hMpc$ \simba\ runs at $z=2.4$ differ only up to 0.1 dex from those found for the $100 \hMpc$ run (further details in appendix \S~\ref{sec:haloes}). We explicitly verified that varying the mass threshold by 0.1 dex in the \simba\ $100 \hMpc$ run has negligible impact on the results of this work (see appendix \S~\ref{sec:threshold} for more details). Thus, we decided to apply the same mass and luminosity cuts throughout all runs, for a more straightforward comparison of the results. Table~\ref{tab:QSOs} summarises the number of QSOs selected in each run, and how many of such QSOs exhibit each AGN feedback mode specified in \S~\ref{sec:feedback}. 

We also verified that even if we selected QSOs above the luminosity threshold providing the best fit to \cite{White_2012} observations, without any reference to the mass of the host haloes, we would obtain the same sample of QSOs for the \simba\ $100 \hMpc$ run. On the contrary, this method and the combined mass-luminosity criterion described earlier yield different QSO samples in the various $50 \hMpc$ runs, the latter resulting in smaller differences among the optimal luminosity cuts across the various runs and generally resulting in a better match to \cite{White_2012} observations than the former. We therefore adopted the combined mass-luminosity criterion as the fiducial one in this work, given that it seems to be more robust and, as already mentioned, it enables a straightforward comparison with mass-based selection method in other numerical studies. Nonetheless, we verified that even if we constructed the QSO sample by following the simpler criterion the main conclusions of this work would be unchanged (see appendices \S~\ref{sec:haloes} and \S~\ref{sec:threshold}).

\subsection{Generating \lya absorption spectra around QSOs}
\label{sec:skewers}

Once we select QSOs in \simba, we generate \lya mock absorption spectra (``skewers'') at different transverse distances around them. To do this, we first choose the $z$-axis of the simulation as the direction of the LOS. Following \cite{Sorini_2018}, we then select skewers by randomly drawing their transverse distance from QSOs from a log-uniform distribution, and their angular coordinate in the $(x,\,y)$ plane from a uniform distribution. We extract 1000 skewers for every bin of transverse distance, the boundaries of which are the same as in the observations by \cite{Prochaska_2013} and \cite{Font-Ribera_2013}. Skewers are drawn cyclically around all QSOs, ensuring an even distribution around the QSO sample.

We obtain the \hi number density $n_{\rm HI}$ along every skewer in our sample by depositing \simba\ gas particles onto a regular grid along that skewer with a cell width of $10 \, \rm km \, s^{-1}$, by means of the publicly available code \texttt{Pygad}\footnote{\url{https://bitbucket.org/broett/pygad/src/master/}} (Cernetic et al., submitted; see also \citealt{Pygad_code, Pygad}).\footnote{We verified that refining the grid down to a cell width of $5 \, \rm km \, s^{-1}$ would not  change the conclusions of our work.} We remind the reader that the \hi number density is a native field of \simba, which is determined by accounting for photoionisation, collisional ionisation and self-shielding through the relationship between photoionisation rate and hydrogen density found by \cite{Rahmati_2013}. The \lya optical depth $\tau$ is computed by convolving the HI number density with a Voigt profile along the LOS, accounting for redshift space distortions and line broadening due to thermal motion and turbulent velocities of the gas particles (see e.g. \citealt{Meiksin_review} for the full derivation). The \lya flux is then simply obtained through the definition $F=\exp(-\tau)$. \texttt{Pygad} allows us to extract several optical-depth-weighted quantities, such as temperature and LOS velocity.

Prior to simulating \lya flux absorption around QSOs, we extract a sample of 10000 random skewers in the whole simulation box, and follow the standard approach of choosing the value of the UVB such that the mean \lya flux of our sample matches the observations by \cite{Becker_2013_mean}. We then use that value of the UVB to compute the \lya flux absorption spectra around QSOs at the redshift of interest. We repeat this procedure for each run considered in this work. This enables a fair comparison among the results of the various runs, as they will all be consistent with the observed mean \lya flux in the IGM. We report the factor by which we rescaled the \cite{HM12} UVB for each run in Table~\ref{tab:runs}.

We verified that choosing a mean \lya flux off by $1\sigma$ from \cite{Becker_2013_mean} data would not change the main conclusions of this work. Likewise, regulating the UVB in \simba\ to match the more recent but indirect estimates of the mean flux of the IGM by \cite{Walther_2019} would also leave our conclusions unchanged (see appendix \S~\ref{sec:mean_flux}). 

One effect we do not consider is local photoionisation from the QSO itself, i.e. the quasar proximity effect. Our QSO feedback is limited to mechanical feedback on large scales, while X-ray feedback only applies very close to the black hole.  Accounting for the proximity effect introduces a host of other uncertainties and parameter choices that we prefer to avoid for the present, so we defer this to future work. For now, we note that any such local contribution would tend to drive down the \lya\ mean absorption, and hence our predictions might be considered an upper limit, which would be reduced at some level by the proximity effect. Also, unless otherwise indicated, whenever we discuss the effect of AGN feedback we refer to the prescriptions implemented in \simba, which does not include the proximity effect.

\subsubsection{Example skewers from \simba}
\label{sec:skewers_res}

 \begin{figure*}
	\includegraphics[width=\textwidth]{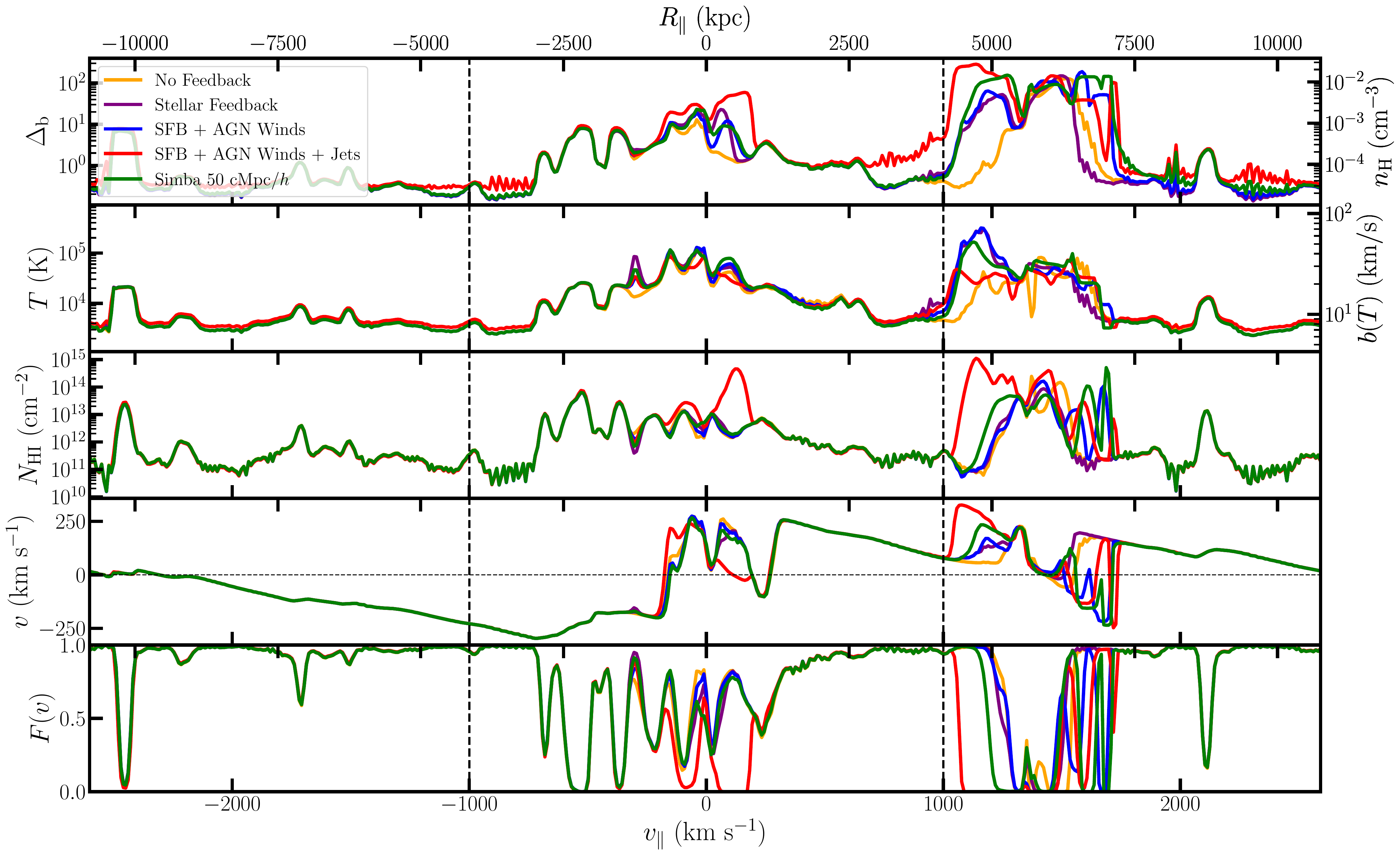}
    \caption{ 
    From top to bottom: gas overdensity (and corresponding total hydrogen density), temperature (and corresponding Doppler broadening), HI column density within a $10 \, \rm km \, s^{-1}$ velocity bin (see Footnote~\ref{foot:NHI} for details), LOS velocity, and \lya transmitted flux along the same skewer located at $120 \kpc$ from the same halo in different runs of the \textsc{Simba} simulation. Orange, purple, blue, red, and green lines refer to the runs with no feedback, stellar feedback only, SFB + AGN winds, SFB + AGN winds + jets, and with all feedback prescriptions active, respectively (see Table~\ref{tab:runs}). The skewer in question spans the whole length of the simulation box. The vertical dashed black lines delimit the velocity window of $\pm 1000 \; \rm km \, s^{-1}$ around the foreground QSO, which is adopted in the measurements by \protect\cite{Prochaska_2013}. Differences on the \lya flux among the various feedback implementations appear only around the most overdense regions.
    }
    \label{fig:skewers}
\end{figure*}
 
Before reproducing the observations of our interest, we visually inspect a sample of skewers generated from the $50 \hMpc$ \simba\ runs. In this way, we can qualitatively assess the impact of the various AGN feedback prescriptions on the simulated \lya spectra.

As an example, in Figure \ref{fig:skewers} we display various physical quantities obtained in the various runs along one skewer throughout the simulation box, located at $\sim 120 \kpc$ from the same QSO host. From top to bottom, we show the optical-depth-weighted gas density and the corresponding total hydrogen density, the optical-depth-weighted temperature and corresponding Doppler broadening, the \hi column density within a $10 \, \rm km s^{-1}$ LOS velocity bin\footnote{ \label{foot:NHI} Note that this is not the operational definition of column density generally adopted in observations, where the column density is usually associated to a line, or ``system''. The quantity that we plot in Figure~\ref{fig:skewers} is $N_{\rm HI}(v_{\parallel}) = n_{\rm HI}(v_{\parallel}) \Delta v_{\parallel}/H(z)$, where $\Delta {v_\parallel}$ is the width of the velocity bin along the LOS centred in $v_{\parallel}$, and $n_{\rm HI}$ is the HI number density.}, the optical-depth-weighted LOS peculiar velocity, and the \lya flux computed as explained in \S~\ref{sec:skewers}. In all panels, the lower $x$-axis reports the redshift-space coordinates in velocity units, relative to the foreground QSO. The upper $x$-axis shows the equivalent coordinates in spatial units, under the assumption of a pure Hubble flow. The vertical dashed lines delimit the $\pm 1000 \, \rm km \, s^{-1}$ velocity window within which we will compute the \lya flux contrast. In all panels, the \simba\ run without feedback is plotted with an orange line, the run with stellar feedback only with a purple line, the run incorporating stellar feedback and AGN winds with a blue line, the run with stellar feedback, AGN winds and jets active with a red line, and the fiducial full AGN feedback run with a green line. In the fourth panel from the top, the horizontal dotted line marks the zero level of the LOS velocity field, to guide the eye.

Overall, the impact of the different feedback prescriptions does not seem to be significant. While stellar feedback and AGN winds have minimal effect on all quantities explored, switching on jets moderately alters the density, LOS velocity and \hi column density skewers in the vicinity of the QSO host, but has more limited impact on the temperature. This is due to the fully kinetic implementation of AGN jets in \simba, which results in an outwards kick to gas particles along the direction of the angular momentum of the BH, without directly injecting heat in the CGM (unlike in \textsc{Illustris}, \citealt{Vogelsberger_2014, Sijacki_2007}). At $z>2$, jets have not been active for enough time to appreciably increase the internal energy (and hence the temperature) of gas surrounding the AGN \citep[see][and the discussion in \S~\ref{sec:discussion}]{Christiansen_2019}. The modifications introduced by the AGN jets in the skewers shown in Figure \ref{fig:skewers} are somewhat compensated by the addition of X-ray heating. However, X-ray heating occurs only within the BH kernel, on scales much smaller than those probed by the skewer shown in Figure~\ref{fig:skewers}. It might well be the case that X-ray heating affects BH growth, possibly reducing the accretion rate and thus the impact of jets. On the other hand, the difference that we observe between the run without X-ray heating and the \simba\ $50 \hMpc$ run may be due to stochastic effects between the two simulations \citep[see, e.g.,][]{Keller_2019}. Therefore, it is hard to establish a precise causal relation between the activation of X-ray heating and the signature on the flux skewer considered.

The general picture appears to be consistent with \cite{Theuns_2002c}, who showed that feedback significantly impacts \lya absorption only around the strongest lines, while leaving flux skewers almost unaffected elsewhere. In our case the differences from run to run around the highest-density regions appear 
to be even somewhat smaller than in their work, perhaps with the exception of the run with AGN winds and jets but not X-ray heating.

Of course, qualitative arguments based on one or few skewers serve only as a tool to develop physical intuition, and should not be used to make conclusive statements. In the next sections, we will investigate the statistical properties of the skewers extracted from the simulations considered in this work, comparing them with observations. This will allow us to gain a deeper understanding of the impact of AGN feedback on the physics of the CGM of $z\sim 2-3$ QSOs.

\section{Results}
\label{sec:results}

In this section we present the results of our work. In \S~\ref{sec:obs} we give an overview of the datasets which we aim to reproduce with the simulations. We then compare observations of the mean \lya flux fluctuations profile around QSOs with the results of the \simba\ $100 \hMpc$ run and the various $50 \hMpc$ runs in \S~\ref{sec:simba_100} and \S~\ref{sec:simba_50}, respectively. 

\begin{figure*}
	\includegraphics[width=\columnwidth]{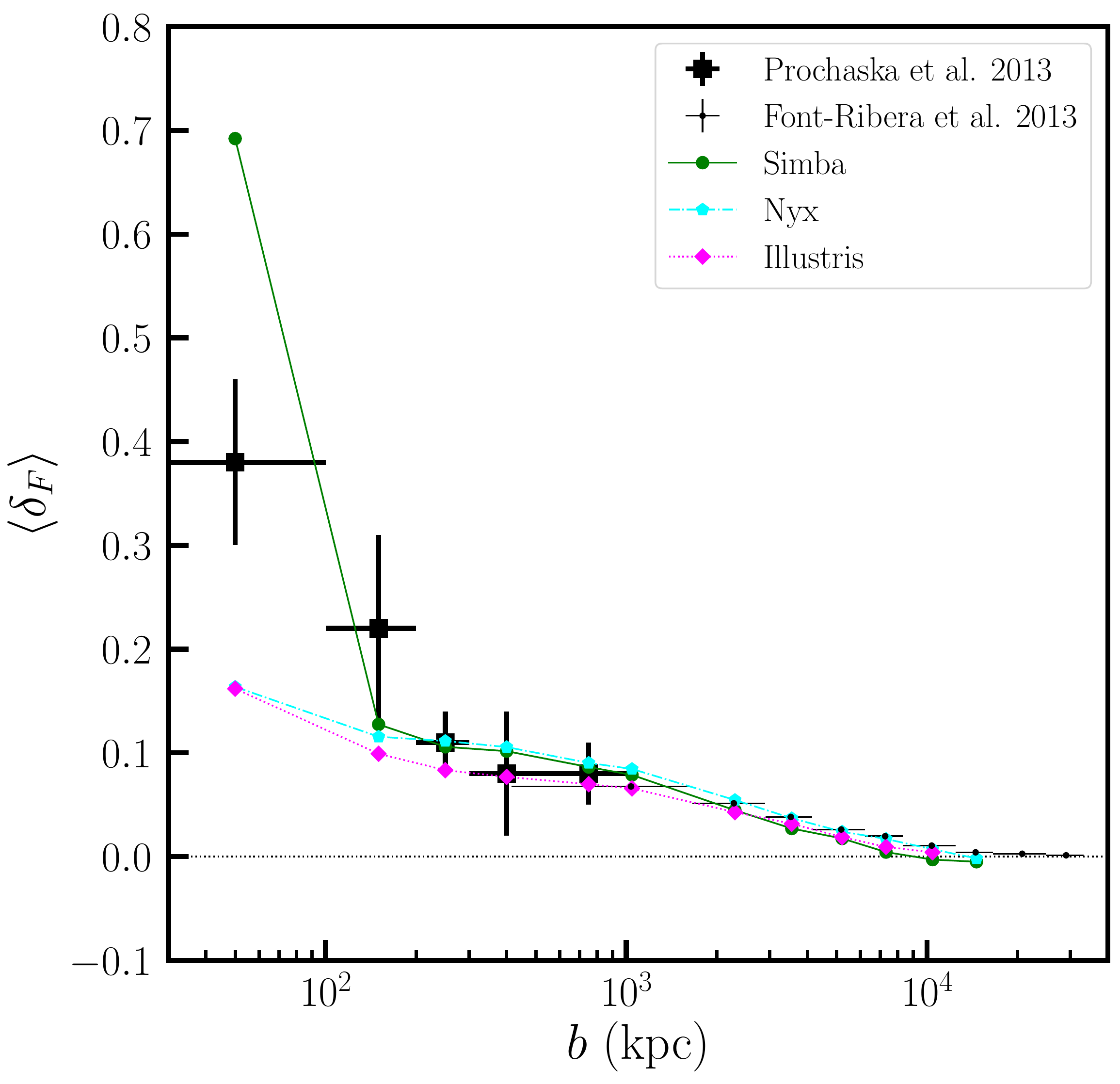}
	\hfill
	\includegraphics[width=\columnwidth]{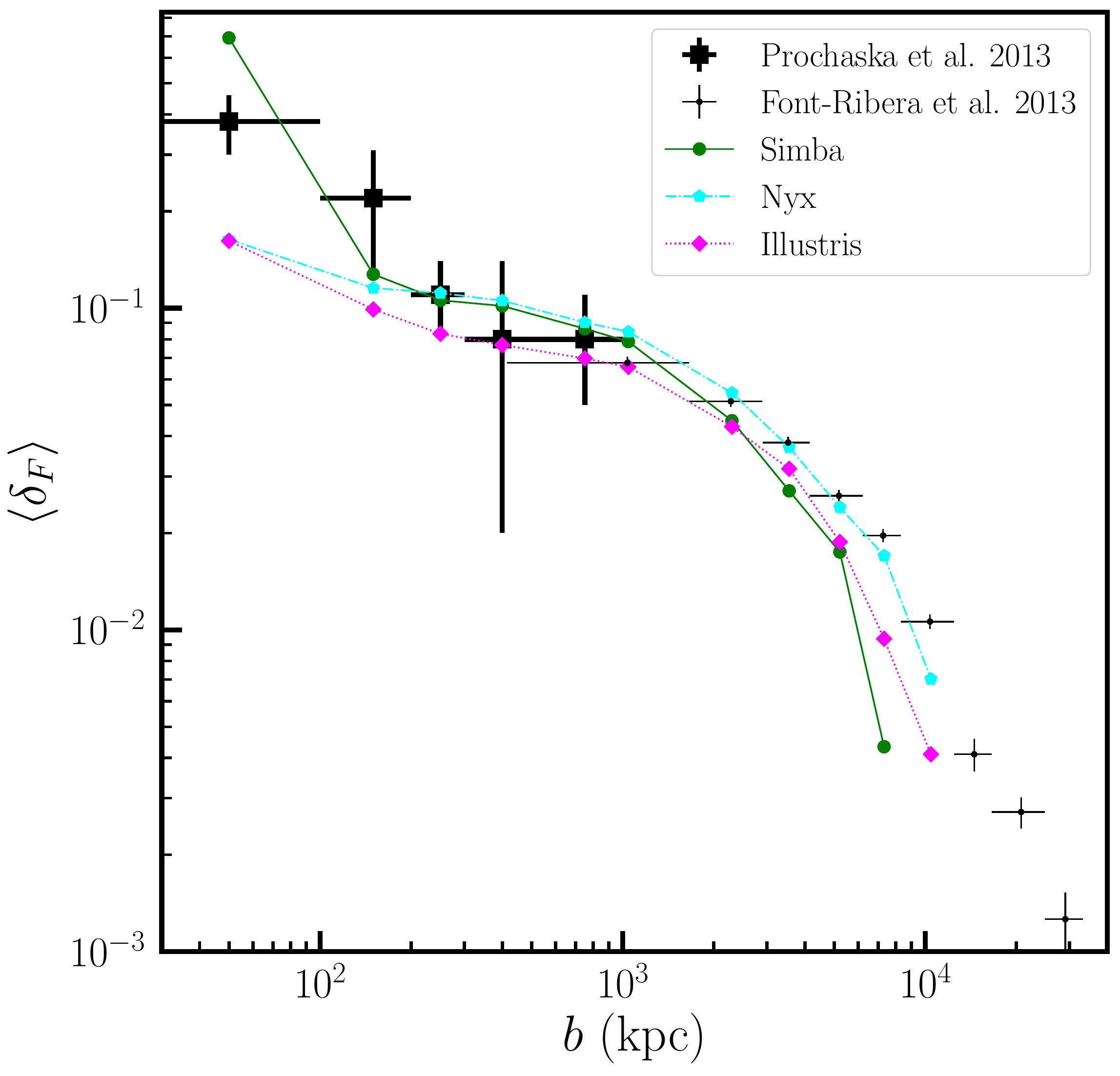}
    \caption{\textit{Left panel}: Mean \lya flux fluctuations profile around foreground QSOs, as a function of their transverse distance from background QSOs. The big square black data points are the measurements by \protect\cite{Prochaska_2013}, while the small round data points show the mean $\delta_F$ inferred from the \lya--QSO cross-correlation measured by BOSS \protect\cite{Font-Ribera_2013}. Vertical bars represent the errors on the measurements, whereas horizontal bars the widths of transverse distance bins. The predictions of the $100 \hMpc$ \simba\ run, the \illustris\ simulation, and the $100 \hMpc$ \nyx\ run considered in \protect\cite{Sorini_2018} are represented as green circles, magenta diamonds, and cyan pentagons, respectively. To guide the eye, all points referring to a certain run are connected with a thin solid line of the same colour. \nyx\ and \illustris\ give the same prediction in the innermost bin (albeit for different physical reasons - see main text for details), though drastically different from \simba. This underscores the importance of \lya absorption measurements within the CGM of QSOs to discriminate among the prescriptions implemented in different simulations. \textit{Right panel}: Same as in the left panel, but with a logarithmic scale on the $y$ axis, to highlight the differences among the predictions of the various simulations. On large scales, Nyx gives the best agreement with observations. However, no simulation is tension with data once the uncertainties within the modelling adopted in this work are taken into account (see main text and Appendix \ref{sec:systematics} for details).}
    \label{fig:deltaF_100}
\end{figure*}

\subsection{Observations}
\label{sec:obs}

Our goal is to compare the results of \simba\ with observations of \lya absorption around QSOs by \cite{Prochaska_2013} and \cite{Font-Ribera_2013}. 

\begin{figure*}
	\includegraphics[width=0.49\textwidth]{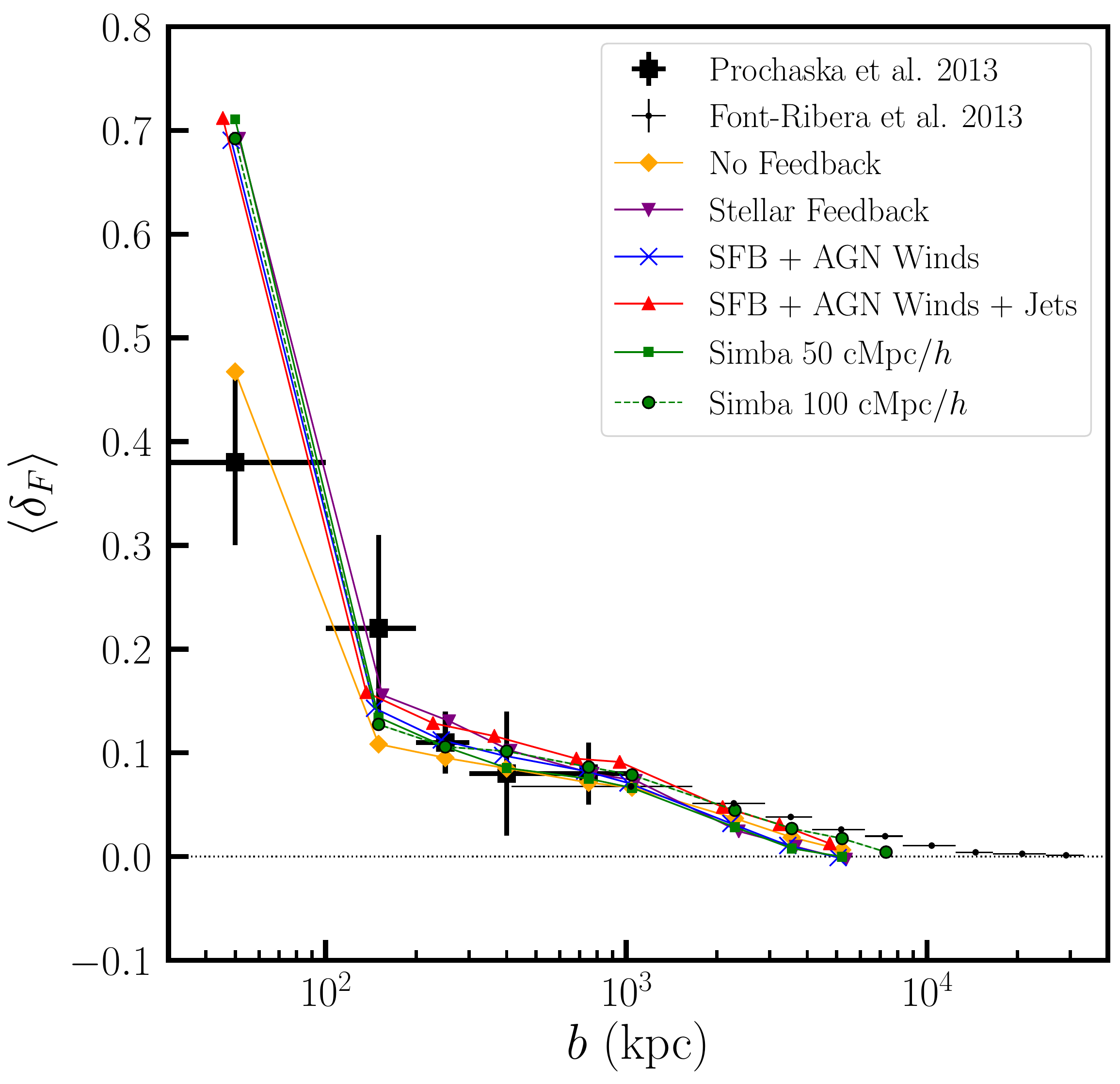}
	\hfill
	\includegraphics[width=0.49\textwidth]{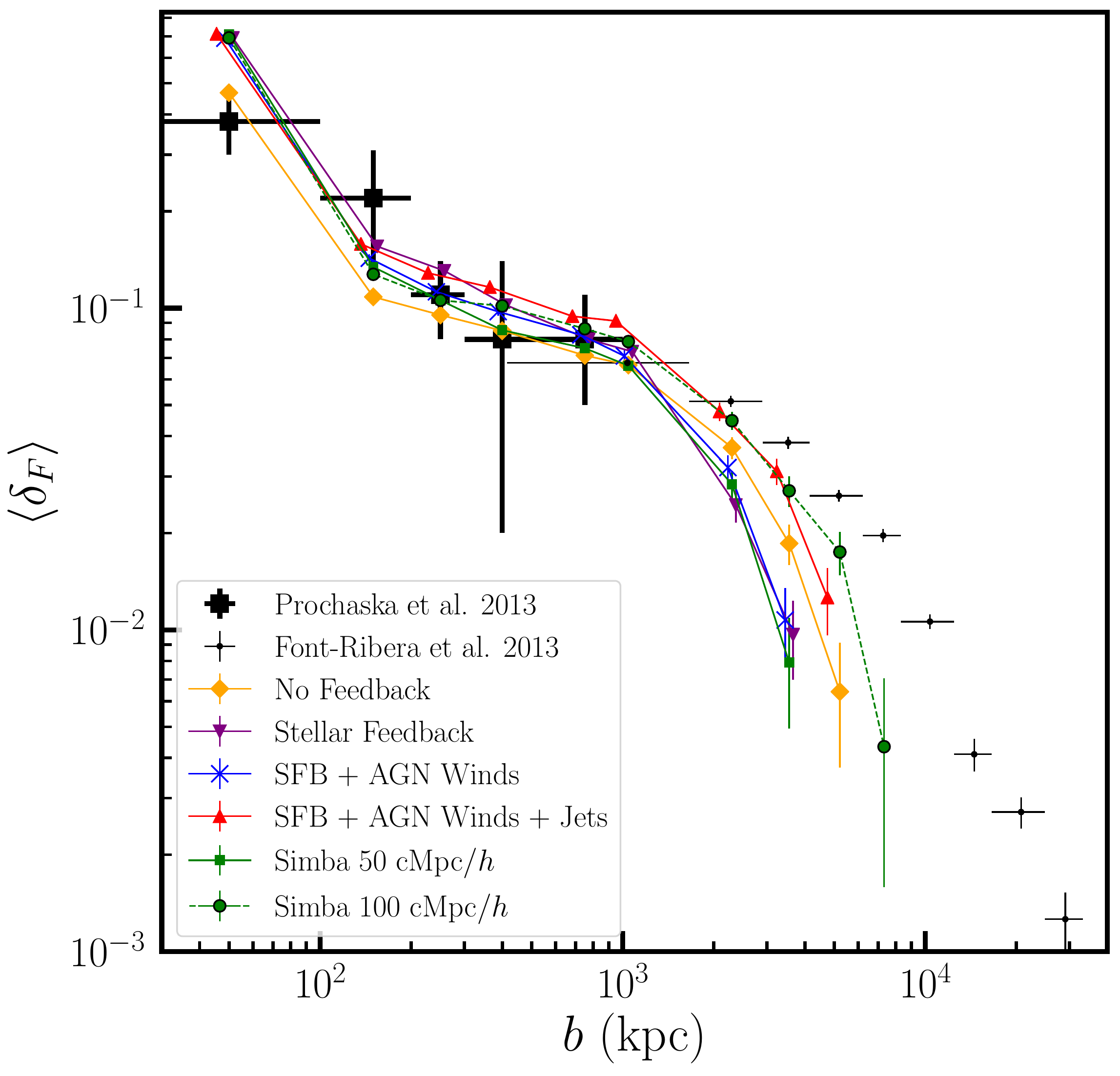}
    \caption{\textit{Left panel}: As in Figure \ref{fig:deltaF_100}, but with a comparison among the various \simba\ $50 \, h^{-1} \, \rm cMpc$ runs at $z=2.4$. Their predictions are displayed as follows: orange diamonds refer to the no-feedback run; purple reversed triangles correspond to the run with stellar feedback; blue crosses refer to the run with  stellar feedback and AGN winds; red triangles represent the run with stellar feedback, AGN winds and jets; green squares are the results of the full stellar and AGN feedback implementation. As a reference, we plot again the results of the \simba\ $100 \, h^{-1} \, \rm cMpc$ run (green circles connected with a thin dashed green line). \textit{Right panel}: Same as in the left panel, but with a logarithmic scale on the $y$ axis. Stellar feedback has the highest impact on the mean \lya flux contrast profile within $100 \kpc$. The error bars on the simulated profiles show the statistical error on $\langle \delta_F \rangle$ due to LOS-to-LOS variance. On larger scales, all runs give similar results. All $50 \, h^{-1} \, \rm cMpc$ runs undershoot BOSS data on scales $b\gtrsim 3 \, \rm Mpc$, highlighting the importance of simulating large volumes in studies of which the primary aim is to reproduce the large-scale \lya--QSO correlations measured with BOSS.}
    \label{fig:delta_F}
\end{figure*}

\cite{Prochaska_2013} observed the spectra of 650 projected QSO pairs in the redshift range $2<z<3$, with transverse separations $< 1\Mpc$. For each background QSO spectrum, they measured the \lya flux contrast within a velocity window of $\Delta v= \pm 1000 \, \rm km \, s^{-1}$, centred around the LOS redshift-space position of the foreground QSO. This quantity is defined as
\begin{equation} 
	\delta_{F} = 1 - \frac{\langle F \rangle _{\Delta v}}{\bar{F}_{\rm IGM}},
\end{equation} 
where $\langle F \rangle _{\Delta v}$ is the mean \lya flux within the aforementioned velocity window, and $\bar{F}_{\rm IGM}$ is the mean \lya flux in the IGM at the same redshift of the foreground QSO. \cite{Prochaska_2013} then grouped the spectra of all QSOs in five bins of transverse distance, and obtained the mean \lya flux contrast $\langle\delta_{F} \rangle$ averaged over all QSOs in each bin. The resulting $\langle \delta_{F} \rangle$ profile as a function of the transverse distance between QSO pairs are reported in Figures \ref{fig:deltaF_100} and \ref{fig:delta_F} with big black squares. The vertical bars indicate the $1\sigma$ errors on the measurements, while the horizontal bars show the widths of the transverse distance bins.

The observations by \cite{Font-Ribera_2013} come from the data of the BOSS survey DR9 \citep{Ahn_2012}. From a sample of $\sim 6\times 10^4$ QSOs in the redshift range $2<z<3.5$, they measured the \lya--QSO cross-correlation function in bins of parallel and transverse distance with respect to the LOS. As shown by \cite{Sorini_2018}, this observable can be converted into a $\langle \delta_{F} \rangle$ profile \textit{a la} \cite{Prochaska_2013}. Within very mild assumptions, the mean \lya flux contrast in a given bin of transverse distance is simply the opposite of the average of the \lya--QSO cross-correlation over the LOS bins falling into the $\pm 1000 \, \rm km \, s^{-1}$ velocity window, weighted by the bin widths along the LOS (we refer the interested reader to the appendix D in \citealt{Sorini_2018} for the full derivation). In this way, despite coming from very different observations, the measurements by \cite{Prochaska_2013} and \cite{Font-Ribera_2013} can be easily compared to each other, and also with theoretical predictions of the mean \lya flux profile.

We show the resulting $\langle \delta_{F} \rangle$ profile obtained from \cite{Font-Ribera_2013} data by \cite{Sorini_2018} with small black circles in Figures \ref{fig:deltaF_100} and \ref{fig:delta_F}. Also for this dataset, the horizontal bars represent the transverse bin widths, while the vertical bars the $1\sigma$ error of the measurements. These are much smaller than in \cite{Prochaska_2013} mainly because of the $\sim 100$ times larger QSO sample. Remarkably, the two datasets are consistent with each other (see in particular the bins at $b\sim 1\Mpc$), and they have the potential to jointly constrain the physics of IGM and CGM over three decades in distance.

\subsection{Mean \lya flux contrast profile in \simba}
\label{sec:simba_100}

We begin with comparing the results of the \simba\ $100 \hMpc$ run with the observations described in \S~\ref{sec:obs}. In the left panel of Figure \ref{fig:deltaF_100} we plot the predicted mean \lya flux contrast profile around QSOs at the median redshift of the observations ($z\approx 2.4$) with green circles, connected with a solid line to guide the eye. We also plot the results obtained with \illustris\ and \nyx\ by \cite{Sorini_2018} with magenta diamonds connected by a dotted line and with cyan pentagons linked with a dashed line, respectively. In the right panel, we show the results of the exact same data sets and simulations on a logarithmic scale, to facilitate the comparison between observations and simulations on the largest scales.

We find that \simba\ is in overall good agreement with observations, albeit \cite{Font-Ribera_2013} data are undershot by the simulation on large scales ($b \gtrsim 2 \Mpc$). However, when taking into account uncertainties in our modelling stemming from the selection of QSO hosts, and from the simplification of extracting skewers only from the snapshot corresponding to the median redshift of the observations (hence, neglecting the actual redshift distribution of foreground QSOs), the predictions of \simba\ are consistent with \cite{Font-Ribera_2013} measurements (see appendix \ref{sec:systematics} for further details). Nevertheless, \illustris\ and even more so \nyx\ provide a better match to the observations on scales $b\gtrsim 4 \Mpc$. 

On intermediate scales ($100 \, \mathrm{\kpc}\gtrsim b \gtrsim 2 \Mpc$) \simba\ and \nyx\ predict the same mean \lya flux contrast profile. Given that \nyx\ does not include any feedback implementations, this implies that in \simba\ the impact of stellar and AGN feedback on $\langle \delta_F \rangle$ is confined within a transverse distance of $100 \kpc$. On the contrary, the gas heating due to the radio-mode AGN feedback in \illustris\ extends out to 3-4 virial radii from QSOs, affecting the $\langle \delta_F \rangle$ profile out to $700-1000 \kpc$ from the foreground object (\citealt{Sorini_2018}; see also \citealt{Gurvich_2017}). 

Within the innermost bin of transverse distance ($b<100 \kpc$) we find quite a different situation. Whereas \nyx\ and \illustris\ give the same result for $\langle \delta_F \rangle$, underestimating the \cite{Prochaska_2013} data point by almost $3 \sigma$, \simba\ drastically differs from the other simulations, overshooting the observations by $\sim 3.5 \sigma$. While this level of tension with data certainly confirms how challenging it is to reproduce the CGM properties within $100 \kpc$ from QSOs, it is perhaps not surprising considering the uncertainties underlying our modelling (see \S~\ref{sec:systematics}), such as any potential transverse proximity effect from the QSO, which would tend to lower the simulated $\delta_F$.

Another factor that could improve the agreement with data within the innermost bin of transverse distance is the inclusion of the error on the redshift of foreground QSOs. In \cite{Prochaska_2013}, the typical error is $\sigma_z=520 \, \rm km \, s^{-1}$. To account for this, we followed the approach adopted by \cite{Prochaska_2013} when comparing their data with simulations, and added a scatter to the LOS-velocity of the QSO hosts in \simba, drawn from a Gaussian distribution with $\sigma_z=520 \, \rm km \, s^{-1}$ \citep[see also][]{Meiksin_2017, Sorini_2018}. We found that introducing such scatter would lower the \simba\ mean \lya flux contrast shown in Figure~\ref{fig:deltaF_100} by $\sim 0.05$ in the innermost bin, and by $\lesssim 0.02$ in all other bins. Variations of this order cannot account for the discrepancies between \simba\ and the other simulations within $100 \kpc$ from QSOs. This is not surprising, given that the large width of the velocity window in the observations ($2000 \, \rm km \, s^{-1}$) is able to mitigate redshift errors of even several hundreds of $\rm km \, \rm s^{-1}$ \citep{Prochaska_2013}.

However, it was also shown that different absorption lines used to estimate the redshift of the QSOs can exhibit offsets up to $1000 \, \rm km\, s^{-1}$ in some cases \citep[see e.g.][]{Paris_2018}. When comparing models to their data, \cite{Font-Ribera_2013} explicitly introduced Gaussian-distributed redshift offsets and dispersions in their modelling. Given that we take all QSOs at the median redshift of the observations, we applied a redshift offset of $-115 \, \rm km \, s^{-1}$ and a dispersion of $450 \, \rm km \, s^{-1}$, which are the values used by \cite{Font-Ribera_2013} for their mid-redshift sub-sample ($2.25<z<2.5$). Also in this case, we find that the resulting mean \lya\ flux contrast profile would be overall shifted towards lower values, by an amount of the order of that obtained following \cite{Prochaska_2013} approach. Once again, this is not a negligible shift, but it cannot account for the discrepancies among the different simulations considered here.

Finally, we also assessed the scatter due to sample variance for \simba. We split the box into eight octants of equal volume, and computed the scatter of the $\langle \delta_F \rangle$ profile in all transverse distance bins. We find $\sim 0.07$ for $b< 100 \kpc$ and $\lesssim 0.02$ otherwise. As such, \simba\ seem to be consistent with the other simulations on large scales, within sample variance, LOS-to-LOS variance (see \S~\ref{sec:simba_50} and the other uncertainties arising from our modelling (see appendix~\ref{sec:systematics}).

In conclusion, the discrepancies among the simulations shown in Figure~\ref{fig:deltaF_100} for $b<100 \kpc$ are genuine, and merit further attention. Upcoming large scale surveys such as WEAVE and DESI are expected to detect more QSO pairs in the redshift range considered here, and will therefore allow for more precise measurements of $\langle \delta_F \rangle$ close to QSOs. Furthermore, instruments such as VLT-MUSE have proven to have a great potential in this respect, being be able to resolve AGN pairs with a transverse separation of $\sim 20 \kpc$ \citep{Husemann_2018}. With smaller error bars in the transverse distance range $0 \, \mathrm{\kpc}<b<100 \kpc$, we will be able to discriminate among the predictions of \nyx, and \illustris\ and \simba. Thus, the mean \lya flux contrast profile confirms to be a potentially powerful tool to constrain simulations. This motivates us to further analyse the detailed impact of the various physical processes implemented in \simba, by investigating the predictions of the various $50 \hMpc$ for the $\langle \delta_F \rangle$ profile. We do this next.

\subsection{Impact of feedback}
\label{sec:simba_50}

In the left panel of Figure \ref{fig:delta_F} we show the predictions of the $50 \hMpc$ \simba\ runs with different feedback prescriptions, compared to the observations by \citealt{Prochaska_2013} (big black squares) and \citealt{Font-Ribera_2013} (small black circles). The meaning of the error bars are the same as in Figure \ref{fig:delta_F}. The results of the various runs are plotted as follows: orange diamonds are the results of the no-feedback run; purple reversed triangles correspond to the run with stellar feedback only; blue crosses refer to the run with stellar feedback and AGN winds; red triangles represent the run with stellar feedback, AGN winds and jets; green squares are the results of the \simba\ $50 \hMpc$ simulation. All points are linked with a thin solid line of the same colour, to guide the eye. We also show again the results of the \simba\ $100 \hMpc$ run (green circles connected by a dashed green line) for comparison. The right panel of Figure \ref{fig:deltaF_100} reports exactly the same data and numerical results, but on a logarithmic scale for the $y$-axis. 

The statistical error on $\langle \delta_F \rangle$ due to LOS-to-LOS variance is $\sim 0.01$ in the innermost bin, and $\sim 0.003$ in the other bins. It is shown with error bars around the simulated profiles in the right panel of Figure~\ref{fig:delta_F}. For $b\lesssim 1\Mpc$ the error bars are smaller than the marker size both on a linear and logarithmic scale. We repeated the tests described in \S~\ref{sec:simba_100} to assess the impact of the error on the redshift of the foreground QSOs in the observations to which we compare the various $50 \hMpc$ runs, and found analogous results. We also estimated the scatter on $\langle \delta_F \rangle$ due to sample variance as in \S~\ref{sec:simba_100}, and found that it amounts to $\sim 0.08-0.12$ for $b<100 \kpc$ and $\lesssim 0.05$ otherwise.

On large scales, the mean \lya flux contrast profile predicted by all simulations converges to the mean \lya flux of the IGM already at $b\approx5 \Mpc$, underpredicting \cite{Font-Ribera_2013} observations. This is a box-size effect, as indicated by the fact that the larger \simba\ $100 \hMpc$ run exhibits a better match with BOSS data and converges to the mean \lya flux of the IGM on larger scales. In fact, $100 \hMpc$ appears to be still too small to fully reproduce all BOSS data points, and we found that also \nyx\ and \illustris\ undershoot \cite{Font-Ribera_2013} measurements, albeit to different extents (see Figure \ref{fig:deltaF_100}). By construction, simulations with different box sizes cannot converge in $\langle \delta_F \rangle$ on the largest scales, and simulations with a box size smaller than the volume required to reproduce the BOSS observations cannot reproduce the full extent of \cite{Font-Ribera_2013} measurements. As such, while for completeness we compare our simulations to the full dynamic range probed by BOSS observations, we cannot exploit the constraining power of the data points on the largest scales. On the other hand, the large-scale regime probed by the $\langle \delta_F \rangle$ profile is not the main focus of this work. We leave an in-depth quantitative comparison between large-scale BOSS data and hydrodynamic simulations with sufficiently large boxes \citep[e.g., IllustrisTNG300, which has a box size of $205 \hMpc$; see][]{Springel_2018} for future work. At the current stage, we limit ourselves to a more qualitative comparison between simulations and data at $b\geq 1 \Mpc$, noting that all runs are comparable with BOSS observations up to $\sim 3 \Mpc$ once the uncertainties inherent to our modelling are taken into account (see appendix \S~\ref{sec:systematics} for more details).

On scales $b\leq 1 \Mpc$, the $50 \hMpc$ and $100 \hMpc$ runs with the full AGN feedback implementation give very similar predictions, meaning that the predictions of the simulations are converged volume-wise in this regime, which represents the main focus of this work.
For $100 \, \mathrm{kpc} \lesssim b \lesssim 1 \Mpc$ , all $50 \hMpc$ runs predict comparable flux contrast profiles, and \cite{Prochaska_2013} data are overall well reproduced by the simulations. The no-feedback run gives the best match to the data in the innermost bin, whereby all other simulations predict essentially the same mean \lya flux contrast of $\sim 0.7$, overshooting the observations. However, this does not mean that the no-feedback run represents a realistic description of the physics regulating galaxy formation. Indeed, it is well known that both stellar and AGN feedback are necessary to reproduce most observables of interest for galaxy formation \citep{Husemann_meeting}; in the case of \simba, \cite{Simba} showed that the inclusion of AGN jets is essential to reproduce the observed stellar mass function. What we do learn from this comparison is that stellar feedback appears to be the dominant driver in determining the average absorption properties of the CGM of $\sim 2-3$ QSOs, with AGN feedback playing a negligible role instead. This is a prediction of \simba, and we will discuss its implications for the physics of the CGM in \S~\ref{sec:discussion}. 

It is still curious that the no-feedback \simba\ run appears to yield a better match to the observations within $100 \kpc$ than the other runs, though. However, we remind the reader that all \simba\ runs do not include radiative feedback from the nearby QSO. Accounting for QSO proximity effects would likely reduce \lya absorption, hence improving the agreement of the fiducial run with the data.

We also point out that if instead of measuring the transverse distance of skewers from the galaxy hosting the QSO in \simba\ we do that by starting from the centre of the host halo, the agreement of all runs with \cite{Prochaska_2013} improves. While this choice is less physically motivated, it is the only viable option in simulations that do not include galaxy formation physics. This was the case of e.g. the \nyx\ run used by \cite{Sorini_2018}, who applied the same criterion for measuring transverse distances from QSO hosts also in \illustris, for consistency. If we adopt the same convention in the \simba~runs, we obtain a better agreement with \cite{Prochaska_2013} in the innermost bin with respect to both \nyx\ and \illustris. We refer the interested reader to \S~\ref{sec:position} for an in-depth discussion.

\section{Properties of CGM/IGM around QSO\lowercase{s} in \simba}
\label{sec:discussion}

In the previous section, we showed how different simulations (\simba, \nyx, and \illustris) can predict very different values for the mean \lya flux contrast within $100 \kpc$ from QSOs. At the same time, we also highlighted that the results from the \simba\ suite of simulations suggest that stellar feedback plays a primary role in the observed absorption properties in the CGM and IGM surrounding QSOs, while the impact of AGN feedback would be marginal. In this section, we want to investigate how feedback processes impact the physical properties of such gaseous media.

\subsection{Radial profiles}
\label{sec:profiles}

\begin{figure*}
	\includegraphics[width=\textwidth]{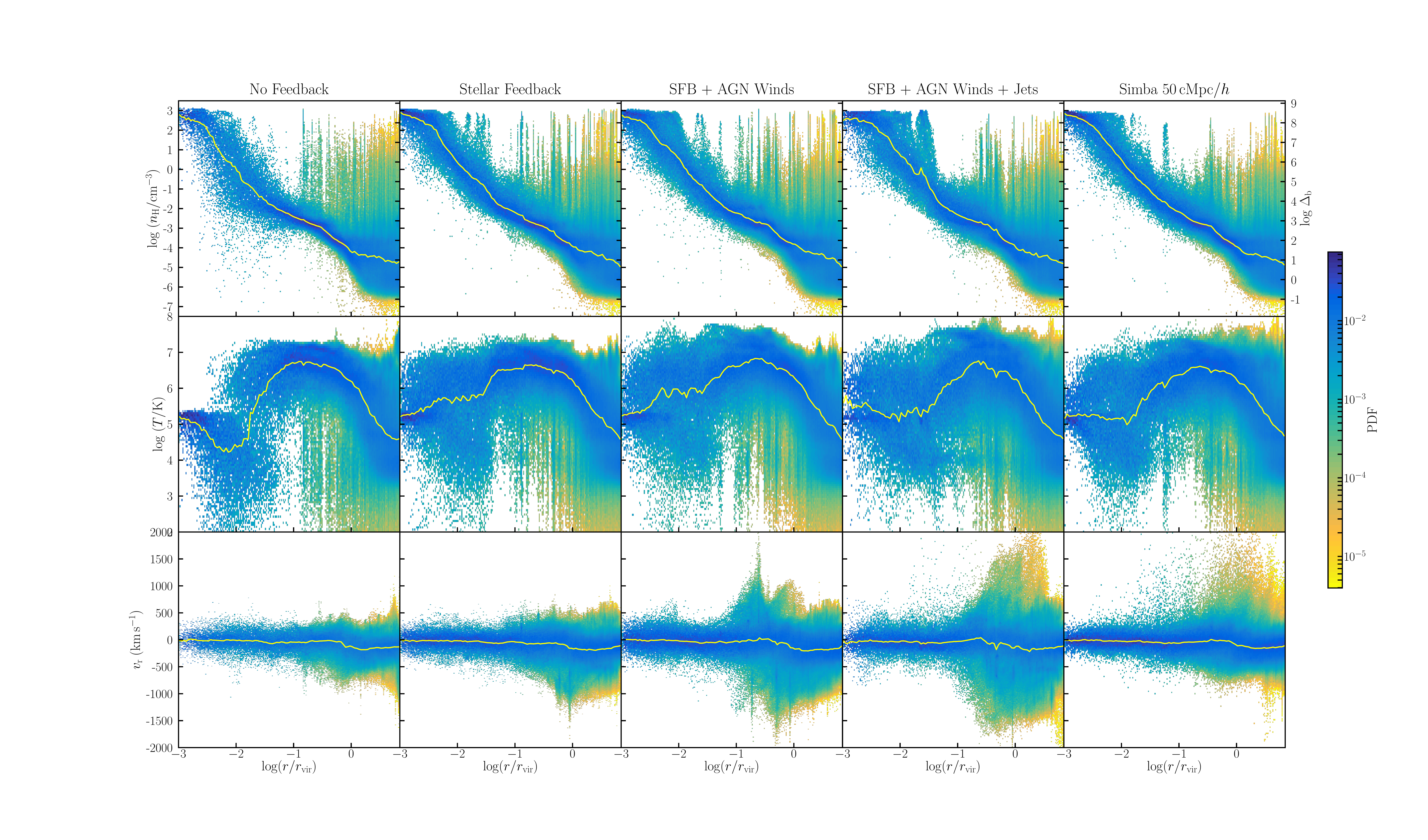}
    \caption{\textit{Top panels}: Radial hydrogen density profile around QSOs in the different $50 \hMpc$ \simba\ runs. All gas particles within $1 \cMpc$ from the centre of all QSO hosts in \textsc{Simba} have been organised in a 2D $n_{\rm H}$-radial distance histogram; for each bin of radial distance, the color bar shows the PDF of hydrogen density for the gas particles within said bin. The yellow line in each panel is the median hydrogen density density profile for the corresponding \simba\ run. The ancillary $y$-axis shows the corresponding gas overdensity. \textit{Mid panels}: As in the top panels, but with the radial profile of the gas temperature. In this case, the colour coding refers to the PDF of temperature for the gas particles in each bin. \textit{Bottom panels}: As in the top and mid panels, but for the radial velocity. For this row, the colour bar corresponds to the PDF of the radial velocity of gas particles in any bin.}\label{fig:r_profiles}
\end{figure*}

\begin{figure}
	\includegraphics[width=\columnwidth]{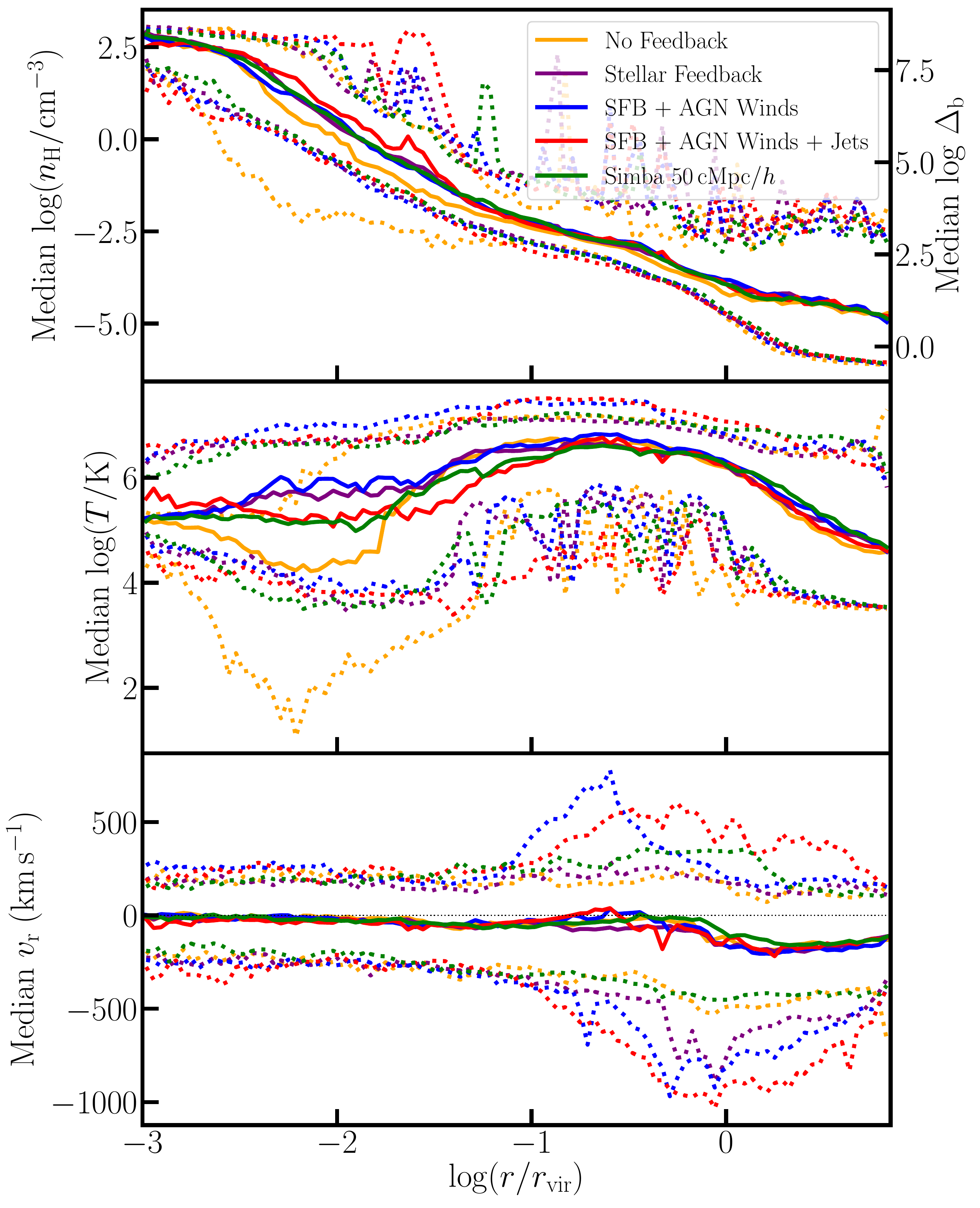}
    \caption{\textit{Top panel}: Median hydrogen number density for all $50 \: h^{-1} \: \rm cMpc$ \simba\ runs, color coded as in Figure \ref{fig:delta_F}. The ancillary $y$-axis shows the corresponding gas overdensity. The dotted lines with the same colour coding mark the 5$^{\rm th}$-95$^{\rm th}$ percentiles of the hydrogen number density PDF within each radial bin. \textit{Mid panel}: As in the top panel, but for the median temperature profile. \textit{Bottom panel}: As in the top and mid panels, but for the median radial velocity profile. Stellar feedback affects mostly the median and spread of the temperature profile, and to a lesser extent, of the density profile, within $\sim 0.1 \, r_{\rm vir}$. Jets from AGN feedback impact the spread of radial velocity profile on scales $\gtrsim 0.1 \, r_{\rm vir}$. These trends are consistent with the results for the mean \lya flux contrast profiles.}
    \label{fig:medians}
\end{figure}

\lya absorption is determined by the local HI number density and temperature, and the peculiar velocity along the LOS. We therefore begin by analysing the radial profiles of three closely related quantities around the QSO samples extracted from our suite of \simba\ simulations. 

For any given  $50 \hMpc$  run, we collect the gas particles within $1 \Mpc$ from all QSOs, and organise them into 100 evenly spaced logarithmic bins of radial distance, normalised to the virial radius of each halo. We then compute the PDF of the density, temperature, and radial velocity of the gas particles falling within every bin of radial distance. The resulting diagrams are shown in the top-row, mid-row, and bottom-row panels of Figure \ref{fig:r_profiles}, respectively. In all panels, the colour bar represents the PDF of the property indicated on the $y$-axis, in any bin of radial distance. The yellow lines are obtained by connecting the medians of the PDFs within all radial bins. As such, from the top to the bottom rows, they represent the median profiles of gas density, temperature and radial velocity, respectively. Every panel along each row shows the results from a different \simba\ run, as specified in the headings at the top of the figure. The ticks on the left and right $y$-axis in the top-row panels show the total hydrogen number density $n_{\rm H}$ and the corresponding gas overdensity with respect to the mean baryon density, respectively. The radial velocity (third row from the top) of gas particles is defined with respect to the centre of the galaxy acting as QSO host, and are defined positive if directed outwards. Although observationally the component of peculiar velocities along the LOS is the one that directly impacts \lya absorption, we chose to analyse the radial velocity because it can provide us with greater physical insight on outflows and inflows within the CGM, while still be related to the LOS velocity. 

The virial radii of the QSOs considered across the various runs fall in the range $95-250 \kpc$. Thus, the softening length ($0.5 \, h^{-1} \, \rm ckpc \approx 0.22 \, \rm kpc$) corresponds to $\lesssim 0.002 r_{\rm vir}$, which lies at the low-end of the $x$-axis in all panels of Figure~\ref{fig:r_profiles}. In the remainder of this section, we will interpret Figure~\ref{fig:r_profiles} by focussing mainly on the range $r/r_{\rm vir} \gtrsim 0.01$, corresponding to $\gtrsim 5-12$ times the softening length. As such, we do not expect the resolution of our simulations to affect our main conclusions.

In all \simba\ models, the median hydrogen density obviously increases moving closer to the QSO. We notice that the median hydrogen density profiles exhibit minimal differences across the different \simba\ runs. This is highlighted in the top panel of Figure \ref{fig:medians}, where we show all median profiles in the same plot, represented by solid lines with the same colour-coding for the different models as in Figure \ref{fig:delta_F}. For every run considered, we also mark the $5^{\rm th}$ and $95^{\rm th}$ percentile of the radial distribution with dotted lines, always adhering to the same colour coding. We can clearly see that both the median and spread of the radial density profile is basically the same for all runs except for the one without any feedback prescription. However, the no-feedback run exhibits a more extended tail towards lower densities only for $r<0.1 \, r_{\rm vir}$, i.e. on galactic scales. We thus conclude that both stellar and AGN feedback appear to have almost no effect on the gas density distribution in the CGM and CGM/IGM interface around QSOs at $z=2.4$ in the \simba\ simulation.

The median temperature increases as we approach the QSOs, but drops in the innermost regions, where the gas is overall cooler and can trigger star formation. Comparing the median profiles and the $5^{\rm th}-95^{\rm th}$ percentiles in the mid-panel of Figure~\ref{fig:medians}, we notice that the no-feedback run is characterised by a dip in the median temperature at $r \sim 0.01 \, r_{\rm vir}$. This feature vanishes when stellar feedback is turned on, because supernovae-driven winds transfer kinetic energy into the surrounding gas. Also, the spread around the median temperature profile becomes symmetric, and not skewed towards lower temperatures as it is the case in the no-feedback run. The excess of gas with temperature $T< 10^4 \, \rm K$ in the no-feedback run is probably due to the increased metal cooling with respect to the stellar feedback run. Switching on AGN feedback modes does not change the median radial profile of gas temperature, nor the spread around the median, as significantly. Therefore, while stellar feedback plays a key role in adding thermal energy to the core of the halo, the impact of AGN winds, jets and X-ray heating on the temperature of the gas is secondary.

The median radial velocity profiles appear to be fairly flat beyond one virial radius. The profiles are slightly negative for $r>r_{\rm vir}$, meaning that there is overall more inflowing than outflowing gas across the QSO sample in all \simba\ runs. Also for the median velocity profiles we do not observe any significant difference across the various models, as highlighted by the bottom panel of Figure \ref{fig:medians}. On the contrary, we do find differences in the spread of the radial velocity distribution around the median. While stellar feedback has little impact if compared to the no-feedback run, the spread around the median stretches up to $\pm 2000 \, \rm km \, s^{-1}$ as AGN jets are introduced (see Figure~\ref{fig:r_profiles}), since they are responsible for a strong injection of momentum in the gas. Though, the signature of jets is actually limited to the increased spread towards positive $v_{\rm r}$. Indeed, the spread in negative velocities is not much larger than that observed in the SFB + AGN winds run, and any differences are likely caused by nearby haloes.

The red dotted lines in the bottom panel of Figure~\ref{fig:medians} tell us that around the virial radius gas particles with radial velocities $\vert v_{\rm r} \vert \gtrsim 1000 \, \mathrm{km} \, \mathrm{s}^{-1}$ (i.e., comparable with or larger than the width of the LOS velocity window in the observations considered in this work) account for $<10\%$ of the total; this represents a generous upper limit to the fraction of such particles beyond $0.1 \, r_{\rm vir}$ from the QSO. Thus, even though the structure of the peculiar velocity field was shown to have a non-negligible impact on the statistical properties of \lya absorption \citep{Sorini_2016}, it does not seem plausible that such a small fraction of outliers could introduce any statistically significant effect on the \lya absorption profile around QSOs. 

In summary, \simba\ shows that stellar feedback is the main actor in determining the physical properties of the gas within the CGM and CGM/IGM interface around QSOs at $z\sim 2-3$. The largest differences in the temperature and density profile occur within $0.1 \, r_{\rm vir}$ (corresponding to $9.5 - 25 \kpc$, depending on the halo within the QSO sample selected in the simulations), and that is reflected in the resulting mean \lya flux contrast. On the other hand, the effect of AGN feedback appears to be important only in shaping the radial velocity profile, but because of the small fraction of the gas particles affected, and their distance from the QSOs, it does not affect the \lya absorption profile appreciably. However, we may still find signatures of the different AGN feedback prescriptions on higher-order statistics, such as the galactocentric temperature-density relationship, which we will investigate in the next subsection.

\subsection{Galactocentric temperature-density relationship}
\label{sec:radial_Trho}

\begin{figure*}
\begin{adjustbox}{addcode={\begin{minipage}{\textheight}}{\caption{%
      Galactocentric temperature-density relationship of the gas surrounding $z=2.4$ QSOs in \simba. Each row corresponds to a different $50 \hMpc$ run; along every row, the first five panels from the left show the temperature-density relationship of gas particles around all AGN hosts, within shells of radial distance progressively farther from the centre of the hosts. The sixth panel from the left shows the temperature-density relationship of the gas particles in the whole volume of the simulation. Stellar feedback has the most visible impact on the galactocentric temperature density relationship, especially within the virial radius, while the different AGN feedback implementations in \simba\ play a marginal role in this respect.
      }\label{fig:radial_Trho}\end{minipage}},rotate=90,left}
     \hspace{0.04\textheight}
     \includegraphics[width=\textheight]{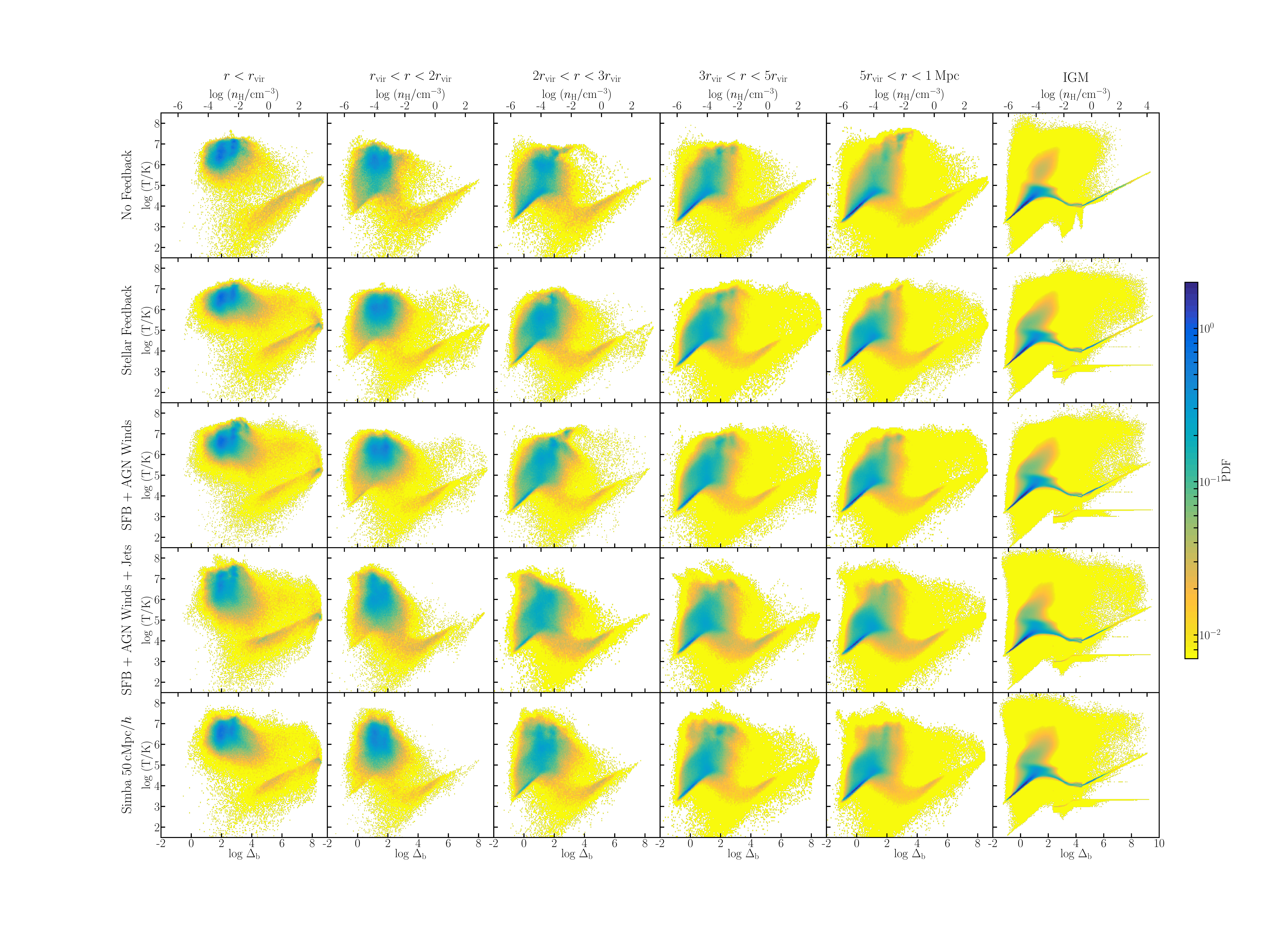}%
\end{adjustbox}
\end{figure*}

The galactocentric temperature-density relationship, i.e. the temperature-density relationship of the gas within different radial shells around the centre of galaxies, is an insightful diagnostic for feedback prescriptions \citep{Sorini_2018}, as it provides information that goes beyond the median properties of the gas.

In Figure \ref{fig:radial_Trho} we show the galactocentric temperature-density relationship of the gas particles around all QSOs. Every row refers to a different $50 \hMpc$ run, as specified on the left side of the figure. Along the same row, the first five panels from left to right report the temperature-density relationship within bins of radial distance extending progressively farther from the QSO. The boundaries of such bins are reported in the headings of the top panels of the figure. The sixth panel from the left shows the temperature-density relationship obtained from all gas particles in the whole simulation box of the corresponding \simba\ run. In all panels, the ticks in the lower $x$-axis refer to the gas overdensity with respect to the mean baryon density, while the ticks in the upper $x$-axis represent the corresponding total hydrogen number density.

The full-box temperature-density relationships look all qualitatively similar across the various runs. They exhibit the characteristic power-law feature of the IGM \citep{Hui_1997} in the density and temperature ranges $10^{-6} \, \mathrm{cm}^{-3} < n_{\rm H} < 10^{-4} \, \rm cm^{-3}$ and $10^3 \, \mathrm{K} \lesssim T \lesssim 10^5 \, \rm K$, respectively. For a quantitative comparison among the different runs, we select the median temperature of the gas particles corresponding to density bins centred in $\log\Delta_{\rm b}=\pm 0.5$ with $5\%$ width, and determine the power law $T=T_0 \Delta_{\rm b}^{\gamma-1}$ connecting the two values of the median temperature. We report the values that we obtained for $T_0$ and $\gamma$ in Table \ref{tab:IGM}. All models converge on the same results, with the run with AGN winds and jets only predicting a slightly higher and flatter relationship. 

Nevertheless, this difference is not that significant if compared to the variations seen across different observations. For instance, \cite{Hiss_2018} found $T_0 = 13334_{-1530}^{+1206}$ and $\gamma = 1.56\pm 0.12$ at $z=2.4$ from Voigt profile fitting of QSO spectra; \cite{Walther_2019} extracted the parameters of the temperature-density relationship from measurements of the \lya forest power spectrum via a Markov Chain Monte Carlo (MCMC), obtaining $(T_0,\, \gamma)=(0.831^{+0.112}_{-0.078}\, \times 10^4\, \mathrm{K},\, 2.07^{+0.13}_{-0.18})$ and  $(T_0,\, \gamma)=(1.165^{+0.29}_{-0.189}\, \times 10^4\, \mathrm{K},\, 1.63^{+0.16}_{-0.19})$, for flat and Gaussian priors on the value of the mean \lya flux, respectively. The value of $\gamma$ found in our \simba\ runs are a very close match to \cite{Hiss_2018}, and consistent within $0.85-2\sigma$ and $0.16-0.4\sigma$ (depending on the run) with \cite{Walther_2019} for the flat and Gaussian priors, respectively. While the $T_0$ predicted by \simba\ is consistent within less than $\sim 0.5\sigma$ and $\sim 2\sigma$ with \cite{Walther_2019} for the flat and Gaussian priors, respectively, it deviates from \cite{Hiss_2018} up to $\sim 3.8-4.5 \sigma$, depending on the run considered. In conclusion, while \simba\ reproduces different measurements of the temperature-density relationship of the IGM within different degrees of accuracy, the values obtained for $T_0$ and $\gamma$ are reasonable given the spread in the observations themselves.

The other power-law feature present in all panels in the last column from the left of Figure \ref{fig:radial_Trho} is a numerical artefact. It stems from the ISM heating prescriptions in \simba, which are activated as gas bound to galaxies overcomes a density thresholds of $0.18 \, \rm cm^{-3}$, at which the temperature is assumed to be $10^4 \, \rm K$ \citep{Mufasa}.

\begin{table}
\begin{center}
\begin{tabular}{lcc}
\hline
Simulation & $\log(T_0/ \rm K)$ & $\gamma$ \\
\hline
No Feedback & 3.90 & 1.60 \\
Stellar Feedback & 3.90 & 1.60 \\
SFB + AGN Winds & 3.90 & 1.60 \\
SFB + AGN Winds + Jets & 3.94 & 1.55 \\
\simba\ $50 \hMpc$ &3.90 & 1.60\\
\hline
\end{tabular}
\caption{Parameters of the power-law temperature-density relationship of the IGM at $z=2.4$ in the $50 \hMpc$ \simba\ runs.}
\label{tab:IGM}
\end{center}
\end{table}

The temperature-density relationship within the virial radius in the no-feedback run exhibits two distinct features, corresponding to the hot and rarefied phase of shock-heated gas ($10^{-4} \, \mathrm{cm}^{-3} \lesssim n_{\rm H} \lesssim 10^{-2}\,\rm cm^{-3}$ and $10^6 \, \mathrm{K} \lesssim T \lesssim 10^7 \, \rm K$), and to the `galaxy phase' corresponding to cold and dense star-forming regions ($T<10^5\,\rm K$ and $\Delta_{\rm b} > 10^4$). The activation of stellar feedback diffuses gas particles, bridging the two regions in the phase diagram. This bridge-like feature appears because supernovae-driven winds heat gas particles in the `galaxy phase', thus moving them upward in the diagram. From the colour coding of the diagram, we can see that at any fixed temperature, the gas density seems to be less skewed towards higher values, consistent with what we already saw in the mid-panel of Figure~\ref{fig:medians}. As a result, there is on average less \lya transmission, and $\langle \delta_F \rangle$ increases. 

Including AGN feedback does not introduce any significant difference in the galactocentric temperature-density relationship. Perhaps the only visible qualitative difference among the \simba\ runs is that, between two and four virial radii, the peak in the gas PDF at $n_{\rm H}\sim 10^{-3} \, \rm cm^{-3}$ and $T\sim 10^6 \, \rm K$ becomes less sharp as AGN jets are turned on. This is probably due to the winds expelling a fraction of the gas particles out of the innermost shock-heated region. 

Moving further away from the QSO, there are progressively less shock-heated gas particles, and more cool and rarefied gas appears. Between two and three virial radii the diagrams begin exhibiting a power-law feature that will eventually give rise to the IGM temperature-density relationship beyond $3\,r_{\rm vir}$. Thus, the CGM/IGM interface lies between $\sim 3\, r_{\rm vir}$ and $\sim 5\, r_{\rm vir}$ from QSOs. 

\subsection{Implications for the physics of gas}

The results presented in \S~\ref{sec:results} show that stellar feedback is the dominant factor in determining the mean $\langle \delta_F \rangle$ profile in \simba, while the impact of AGN feedback is minimal in this respect. Furthermore, the analysis in \S~\ref{sec:profiles}-\S~\ref{sec:radial_Trho} leads to analogous conclusions on the impact of feedback processes on the thermodynamics of gas within $1 \Mpc$ and $0.1 \, r_{\rm vir}$ from $z\sim2-3$ QSOs, respectively. 

One might question the existence of a causal connection between these two results based on the fact that all plots discussed in \S~\ref{sec:discussion} are made by considering the whole sample of QSOs in our \simba\ runs, and not single QSOs. In fact, as we activate any AGN feedback mode, that does not necessarily mean that all QSOs will actually exhibit that specific mode at $z=2.4$. In particular, only one QSO host in the full \simba\ $50 \hMpc$ run is actually affected by all AGN modes (see Table~\ref{tab:QSOs}). Thus one could in principle argue that AGN feedback processes might actually have a stronger impact on the properties of the gas, but that their signatures on the \lya absorption profiles, as well as on stacked radial profiles and galactocentric temperature-density relationships, might be dimmed because of statistical reasons. However, we explicitly verified that even if we focus on the one QSO with all AGN feedback modes in the full \simba\ $50 \hMpc$ run, and on the corresponding QSOs in the other $50 \hMpc$ runs, the results are consistent with Figures~\ref{fig:r_profiles}-\ref{fig:radial_Trho}. 

We therefore conclude that our results on the properties of the CGM around QSOs are physical, and not the result of a statistical fluke. Consequently, the mean \lya flux contrast predicted by the various \simba\ runs simply reflects the physical differences in the underlying properties of the gas. The dominance of stellar feedback over AGN feedback in shaping such properties is thus a genuine prediction of the \simba\ simulation. It is consistent with \cite{Christiansen_2019}, who showed that while at $z=0$ AGN-driven heating pervades almost the entire simulation box \citep[with $\sim 40\%$ of baryons having moved out of their host halo; see][]{Borrow_2020}, the volume fraction of hot gas is smaller at higher redshift. In particular, regions of hot gas seem to be limited within the CGM of AGN hosts at $z=2$. The extent of the heated gas region is thus expected to be even smaller in the redshift range considered in this work ($2 \leq z \leq 3$). 

This result may still look somewhat surprising to some readers, who might question how realistic the implementations of feedback processes are, especially in light of the discrepancy  between \simba\ and the \cite{Prochaska_2013} measurement closest to QSOs (see Figure~\ref{fig:deltaF_100}). In point of fact, we stress that \simba\ has already proven to successfully reproduce several observable properties of galaxies (e.g., the stellar mass function, see \citealt{Simba}) and black holes \citep{Thomas_2019}. Thus, we consider \simba\ feedback prescriptions to be overall physically sensible, and instead argue that the properties of the CGM in the vicinity of $z\sim 2-3$ QSOs are inherently challenging to reproduce for cosmological simulations, being determined by the interplay of several sub-grid physical processes (see also \S~\ref{sec:previous_work}).

As mentioned previously, a potential resolution of this discrepancy between \simba\ and observations within $100 \kpc$ from $z=2.4$ QSOs would be to drop our assumption of a spatially-uniform ionising background even close to QSOs. This transverse proximity effect has been elusive to quantify, but it has certainly been detected \citep{Dobrzycki_1991, Adelberger_2004, Goncalves_2007, Worseck_2007, Kirkman_2008, Schmidt_2017, Jalan_2019}. If some transverse proximity effect were implemented, it would increase the ionised fraction of \hi in the proximity region of QSOs, pushing the predictions of \simba\ towards lower values of $\langle \delta_F \rangle$, thus improving the agreement with data. We will examine this in future work.

We stress that our claims on the role of stellar and AGN feedback with respect to the CGM and CGM/IGM interface around $z\sim 2-3$ QSOs are limited to \simba\ only. Because of the non-trivial interdependence of stellar and AGN feedback \citep{Booth_2013}, it might still be necessary to include some form of AGN feedback in other simulations to explain CGM properties in QSO environs. Our findings should therefore be treated as the result of a ``numerical experiment'' specific to \simba, and our conclusions cannot be automatically extended to the real Universe. Nonetheless, we highlight that if the actual behaviour of the Universe reflects our results, this would have profound implications for our understanding of the physics of the CGM around $z\sim2-3$ QSOs. Indeed, it would mean that the \textit{average} properties of the gas even around the most luminous BHs could be described without any reference to AGN feedback mechanisms such as winds, jets, and X-ray, or at least without any particularly detailed modelling thereof. 

Obviously, if one were to reproduce observations of outflows around a specific QSO \citep[e.g.][]{Husemann_jet}, one may need to include the necessary AGN-driven physics in the theoretical explanation. However, the properties of the gaseous environment of a large enough population of randomly chosen $z\sim 2-3$ QSOs would remain unaffected by any such mechanism, or at least AGN feedback processes would be sub-dominant with respect to stellar feedback.

Clearly, it is essential to pursue studies similar to our own with other simulations. Indeed, should our result be confirmed by very different simulations too (e.g., EAGLE or IllustrisTNG), then it would make our conclusions on the physics of the CGM of $z\sim 2-3$ QSOs more robust. In the opposite case, it would open up a fruitful debate that would eventually improve our understanding of the physics of gas in QSO environs.

\subsection{Comparison with previous work}
\label{sec:previous_work}

In this section, we will discuss our results in the context of relevant literature in the subject, as well as the foreseeable future challenges for the understanding of the physics of the CGM and IGM, from a theoretical and numerical point of view.

Our conclusions are corroborated by the results of the Sherwood suite of hydrodynamic simulations \citep{Bolton_2017, Meiksin_2017}, which show that the inclusion of stellar feedback is essential (and perhaps sufficient) to reproduce the measurements by \cite{Prochaska_2013}. However, AGN feedback was not implemented in Sherwood, therefore it was not possible to assess its effect relative to stellar feedback. 

Other works in the literature focussed on the related covering fraction of Lyman limit systems around QSOs. \cite{Faucher-Giguere_2016} was able to reproduce the observations by \cite{Prochaska_2013_QPQ5} with high-resolution zoom-in FIRE simulations, implementing stellar feedback only. \cite{Rahmati_2015} reproduced such measurements with the EAGLE suite of simulations, the fiducial runs of which include both stellar and AGN feedback. However, the authors also show that while stellar feedback has a significant impact on the covering fraction profile, adding AGN feedback makes hardly any difference. Thus, both FIRE and EAGLE provide results broadly in agreement with our findings, with the caveats that the observable considered in the aforementioned work is not the same as ours, and that the halo mass of the QSOs selected ($10^{11.8} \leq M_{\rm halo} \leq 10^{12.2}$) coincides with the lower end of the mass range in the \simba\ QSO samples. Finally, we note that adaptive mesh refinement (AMR) simulations with stellar feedback only and radiative transfer in post-processing \citep{Ceverino_2010, Ceverino_2012, Dekel_2013} underpredict  \cite{Prochaska_2013_QPQ5} observations of the covering fraction profile \citep{Fumagalli_2014}, thus they are in contrast with the aforementioned literature.

As already mentioned earlier, there is strong tension between the predictions of the fiducial \simba\ run and \illustris\ on the mean \lya flux contrast within $100 \kpc$ from $z\sim 2-3$ QSOs. \illustris\ predicts much more \lya transmission than \simba. This could be partially because the excess of UV radiation from the nearby QSO is taken into account in \illustris, and partially because the \illustris\ radio-mode AGN feedback appears to heat gas out to $3-4 \, r_{\rm vir}$ from the QSO (\citealt{Sorini_2018}; see also \citealt{Gurvich_2017}). Although it seems reasonable that such feedback prescription dominates the heating of the CGM, this should be explicitly verified by comparing different runs of \illustris\ (or rather the upgraded IllustrisTNG simulation, \citealt{IllustrisTNG}) with and without stellar/AGN feedback.

The fact that \nyx\ and \illustris, despite being radically different simulations, give the same predictions in the innermost bin of \cite{Prochaska_2013} observations highlights how challenging it is to interpret observations in the CGM of QSOs. The reason behind this curious result is that \nyx\ generates hotter but denser radial profiles around QSOs if compared to \illustris; these differences impact the amount of \lya absorption in opposite ways, and appear to somewhat coincidentally compensate for each other \citep{Sorini_2018}. In this work, we were also able to link the physics of CGM/IGM around QSOs with the corresponding \lya absorption properties by analysing the radial profiles and the galactocentric temperature-density relationship, confirming the value of such tools to investigate the impact of feedback on the gas in QSO environs.

The no-feedback \simba\ run predicts $\langle \delta_F \rangle \approx 0.47$ in the innermost bin, whereby the \nyx\ feedback-free hydrodynamic code predicts $\langle \delta_F \rangle \approx 0.17$. There is a caveat about this comparison though, because in our work we measure the transverse distance of LOSs from the position of the central galaxy acting as QSO host, and not from the centre of the halo, as \cite{Sorini_2018} did in their analysis with \nyx. If we adopt the same choice for the origin of the LOS distance in the no-feedback run, then we obtain $\langle \delta_F \rangle \approx 0.35$ (see \S~\ref{sec:position} for further discussion). Even in this case, \nyx\ exhibits less absorption than the no-feedback \simba\ run. This is not surprising, as star formation is not implemented in \nyx, and the cooling function assumes primordial abundances. On the other hand, \simba\ does include star formation and metals. As a result, the gas in the innermost regions of galaxies can cool more efficiently in \simba\ than in \nyx, hence producing more \lya absorption. From Figure~\ref{fig:r_profiles} we can indeed see that in the no-feedback \simba\ run the gas can reach temperatures $\lesssim 10^5 \, \rm K$ for $r\lesssim 0.2 r_{\rm vir}$, while the median temperature of the gas in the innermost regions of haloes in \nyx\ can be about one order of magnitude larger (see \citealt{Sorini_2018}).

On top of the extra physics present in the no-feedback \simba\ simulation, there is also a resolution issue to consider when comparing it with \nyx. Specifically, \nyx\ follows the evolution of gas on a regular Cartesian grid, with a cell size of $35.6 \kpc$. This means that the innermost bin of \cite{Prochaska_2013} observations encompasses less than three resolution elements. Therefore, \nyx\ cannot resolve the high-optical depth $< 500 \, \rm pc$ clouds in an otherwise diffuse CGM implied by observations of \lya absorption around foreground $z\approx 2.5$ galaxies (\citealt{Crighton_2015}; also see \citealt{Simcoe_2006} and \citealt{Crighton_2013}). As a result, \nyx\ results in overall less absorption.  This highlights the need for at least moderately good resolution to robustly model the CGM radial \lya\ profile.

In general, it is important to bear in mind that resolving the small-scale structure of the CGM is challenging for all kinds of cosmological simulations, and this is not expected to improve in the foreseeable future. Indeed, the size of high-column density clouds in the aforementioned observations would require a cell size of $\lesssim 140 \, \rm pc$ in AMR simulations and a resolution better than $4 \, M_{\odot}$ in SPH codes (\citealt{Crighton_2015}; see also \citealt{Agertz_2007, Stern_2016, McCourt_2018}). On the other hand, recent zoom-in simulations built upon the moving-mesh code \texttt{Arepo} were able to achieve a uniform resolution within the CGM of $1 \ckpc$ \citep{van_de_Voort_2019}, while zoom-in simulations utilizing AMR codes could resolve even $\sim 500 \, \rm cpc$ scales \citep{Hummels_2019, FOGGIE_1, FOGGIE_2}. A length scale of $500 \, \rm cpc$ corresponds to $\sim 165 \, \rm pc$ at $z\approx 2.4$, which is about the resolution target for AMR codes that \cite{Crighton_2015} argued for. However, \cite{Arrigoni-Battaia_2015} invoked the presence of even smaller clouds ($\lesssim 20 \, \rm pc$) as an explanation for the high surface brightness of extended giant \lya nebulae around QSOs. Such scales appear to be still beyond current resolution limits of even zoom-in simulations for massive halos that would host QSOs.

Due to numerical constraints, the aforementioned resolution requirements will not likely be achieved in the near future for full-box cosmological hydrodynamic simulations. Though, this does not mean that we should relinquish the ambition of achieving a consistent description of the CGM and IGM, from galactic scales out to $\sim 100 \Mpc$. Rather than exclusively relying on technology-driven advancements in computing power to push cosmological simulations to higher and higher resolution, currently the development of more accurate and physically motivated sub-grid models (as is the case for stellar and AGN feedback mechanisms) seems to be a better strategy worth pursuing. For this reason, it is important to exploit the constraining power of as many observables as possible in order to keep improving feedback prescriptions and ultimately succeed in this enterprise. Indeed, it might be the case that even without reproducing the fine structure of the CGM that is supported by the aforementioned observations, it will still be possible to get the global physical picture right.

Obviously, fully understanding the complex physical mechanisms shaping galaxy formation and converging on the right sub-grid models will take time. This work represents one step in this long term-effort. While the main conclusions are corroborated by some literature, it may well be that other cosmological simulations will find different results. In fact, as discussed earlier, both agreement and discrepancies among simulations have already occurred in the past. Any debate will eventually be settled by upcoming observations, which will drive the improvement of simulations and will increase our understanding in the physics of the CGM and IGM.

\section{Conclusions and perspectives}
\label{sec:conclusions}

The purpose of this work is investigating the properties of the CGM and IGM surrounding $z\sim 2-3 $ QSOs, how they are affected by feedback processes, and what the signatures of these physical drivers on the \lya absorption properties of the gas are. We used several runs of the \simba\ cosmological hydrodynamic simulation: one with no feedback, one with stellar feedback only, and others with the addition of different AGN feedback prescriptions. We compare the mean \lya flux contrast profile around $z\sim 2-3$ QSOs measured from observations of QSO pairs \citep{Prochaska_2013} and inferred from the \lya--QSO cross-correlation measured by \cite{Font-Ribera_2013} from BOSS DR9 \citep{Ahn_2012} data with the predictions of our suite of simulations. We hereby summarise our main findings.

\begin{enumerate}
	\item All runs broadly agree with each other, and with the data, over two decades of transverse distance from foreground QSOs ($100 \, \mathrm{kpc} \lesssim b \lesssim 10 \, \rm Mpc$). Within $100 \kpc$, the simulations with at least stellar feedback overpredict the observed mean \lya flux contrast by $\sim 3.5 \, \sigma$ (Figure~\ref{fig:delta_F}). 
	\item Within $100 \kpc$ from the foreground QSO, stellar feedback has the most significant impact on the predicted mean \lya flux contrast, while the impact of all AGN feedback prescriptions is marginal. 
	\item We investigated the physical properties of the gaseous environment surrounding the QSO samples selected in the various \simba\ runs by examining the radial gas density, temperature, and radial velocity profiles out to $1 \Mpc$ from the QSOs (Figures~\ref{fig:r_profiles}-\ref{fig:medians}). We found that stellar feedback primarily impacts the radial temperature profile, and to a lesser extent the gas density profile, within $\sim 0.1 \, r_{\rm vir}$, while leaving the radial velocity profile almost unchanged. The opposite is true for AGN feedback, in particular in the jet mode: the spread of the gas radial velocity increases, particularly outside $\sim 0.1 \, r_{\rm vir}$, while the effect on temperature and density is comparatively lower. 
	\item We also examined the temperature-density diagram of the gas within different radial shells from the centre of the QSO host (`galactocentric temperature-density relationship'; see Figure~\ref{fig:radial_Trho}). While in the no-feedback run the gas is separated into a hot and rarefied phase and a cold and dense `galaxy' phase within the virial radius, stellar feedback gives rise to a larger amount of hot and dense gas. Also in this case, the impact of AGN feedback appears to be minimal.
\end{enumerate}

From these results, the main conclusion of our work is that, according to the physical models implemented in the \simba\ simulations, stellar feedback is the primary physical driver of the average properties of the gas in the CGM and at the CGM/IGM interface surrounding $z\sim 2-3$ QSOs, while the impact of AGN feedback is minimal. The subsequent implication for observations is that, whereas accounting for AGN-driven winds, jets or X-ray heating may be important for the interpretation of spectra around single QSOs, a detailed modelling of these processes may not be necessary when investigating the average properties of gas surrounding a large sample of QSOs. Obviously, this results is specific to \simba, thus it should be investigated with different simulations as well. We also stress that at the current stage \simba\ does not include increased photoionisation from nearby AGN, which may have a more significant signature on the physical state of the CGM than the aforementioned AGN feedback processes, and could probably improve the agreement with the observations of the mean \lya flux contrast within $100 \kpc$ from QSOs.

From a methodological standpoint, we highlight the following remarks:
\begin{enumerate}
 	\item Our selection criterion of QSO hosts in \simba\ guarantees consistency with the observed autocorrelation function of QSOs \citep{White_2012} and with the typical observed luminosities of QSOs, and furthermore allows for a direct comparison with results of previous works adopting a selection method based on the halo mass of the QSO host rather than its accretion rate;
 	\item We tested our results against possible systematics that may affect our selection criterion of QSOs and our procedure to generate flux skewers from the simulation, and verified that none of such systematics would affect the conclusions of our work;
 	\item We re-iterate that analysing radial profiles of thermodynamic and kinematic properties of gas surrounding QSOs in simulations, as well as visualising the galactocentric temperature-density relationship, are exquisite tools for the understanding of gas physics and of the absorption properties in the CGM and at the CGM/IGM interface around QSOs (as already pointed out by \citealt{Sorini_2018}). 
\end{enumerate}

We also compare the predictions of our fiducial $100 \hMpc$ \simba\ run with those of \nyx\ and \illustris\ cosmological simulations, reported by \cite{Sorini_2018}. The mean \lya flux profiles given by all simulations broadly agree with observations for $b \gtrsim 100 \kpc$. Within $100 \kpc$ from the QSO \nyx\ and \illustris\ give similar predictions, while \simba\ results in much larger absorption (Figure~\ref{fig:deltaF_100}). This shows that the mean \lya flux contrast profile has the potential to become a powerful way to constrain simulations. Indeed, while the precision of current observations does not yet enable making fully conclusive statements in this respect, the error bars are expected to shrink in the immediate future owing to the increased number of QSO pairs to be discovered. Instruments such as VLT-MUSE have already proven to be able to detect QSO sources as close as $\sim 20 \kpc$ at $z\sim 3$ \citep[e.g.][]{Husemann_2018}. Furthermore, large-scale surveys such as WEAVE \citep{WEAVE}  and DESI \citep{DESI} promise to increase the overall number of known QSOs by a factor of $\sim 2$, and to collect spectra at higher resolution and signal-to-noise than BOSS, thus increasing the precision of observations.

An immediate perspective of this work would be to repeat our analysis with other state-of-the-art cosmological hydrodynamic simulations, such as IllustrisTNG and EAGLE. Zoom-in simulations would be beneficial for a more detailed study of the effect of stellar/AGN feedback prescriptions within $\sim 1 \Mpc$ from QSOs. Another interesting line of work consists in investigating the effect of feedback on the mean \lya flux profile around other objects, such as LBGs and DLAs \citep{Meiksin_2017, Turner_2017, Sorini_2018}. Measurements of this observable are already available, and others are still ongoing or scheduled in the near future \citep{Font-Ribera_2012b, Turner_2014, Rubin_2015, CLAMATO_pilot, CLAMATO_DR1, WEAVE, DESI, LATIS}.

\section*{Acknowledgements}

We thank the anonymous referee for useful comments and suggestions, which helped improving the quality of this manuscript. We are grateful to Joseph Hennawi for insightful comments on a draft of this manuscript. We also thank Zarija Luki\'c, Andrea Macci\`o, Teresita Suarez, Jos\'e O$\tilde{\rm n}$orbe, Robert Crain, and Rieko Momose for helpful discussions. We acknowledge the \texttt{yt} team for development and support of \texttt{yt}, and Bernhard R\"ottgers for development of \texttt{Pygad}. DS is supported by the European Research Council, under grant no. 670193. RD acknowledges support from the Wolfson Research Merit Award program of the U.K. Royal Society. DAA acknowledges support by the Flatiron Institute, which is supported by the Simons Foundation, and which we thank for the kind hospitality. This work used the DiRAC\MVAt Durham facility managed by the Institute for Computational Cosmology on behalf of the STFC DiRAC HPC Facility. The equipment was funded by BEIS capital funding via STFC capital grants ST/P002293/1, ST/R002371/1 and ST/S002502/1, Durham University and STFC operations grant ST/R000832/1. DiRAC is part of the National e-Infrastructure. 
This work made extensive use of the NASA Astrophysics Data System and of the astro-ph preprint archive at arXiv.org.\\

\noindent
DS dedicates this work to the memory of his grandmother Lucilla, who passed away as this manuscript was being finalised.

\section*{Data availability}

The data underlying this article are available at \url{http://simba.roe.ac.uk}.




\bibliographystyle{mnras}
\bibliography{MN-20-1863-MJ_accepted} 



\appendix

\section{Details on the selection of QSO\lowercase{s}}

\subsection{Optimal mass and luminosity thresholds in \simba}
\label{sec:haloes}
\begin{figure}
	\includegraphics[width=\columnwidth]{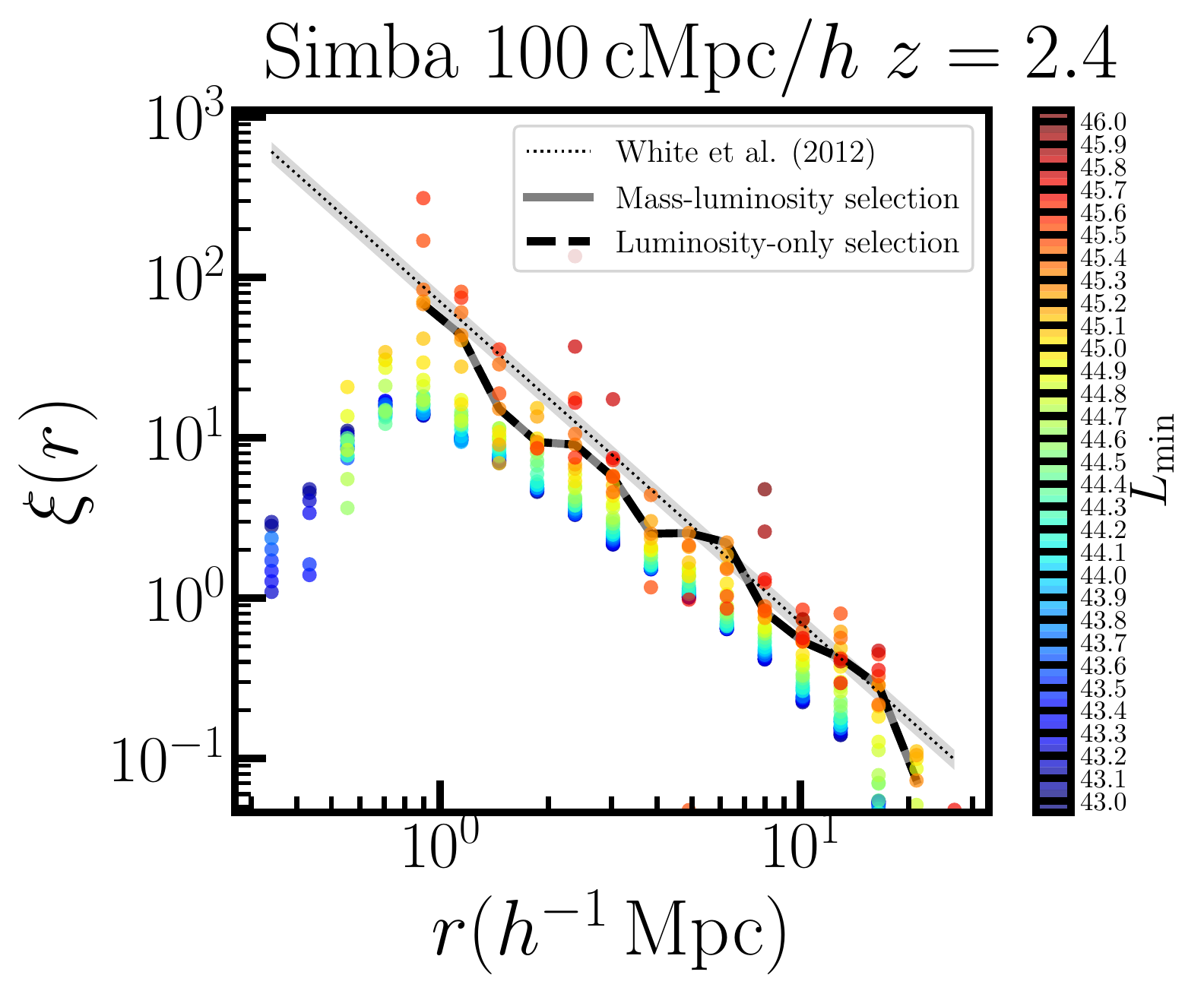}
	\caption{Autocorrelation function of QSOs taken from the $100 \hMpc$ \simba\ run. The coloured points represent the correlation function of QSOs with luminosity above the threshold indicated in the colour bar. The dotted black line is the best-fit power law to the QSO clustering observations \citep{White_2012}. The shaded grey area around such power law indicates the corresponding error within 1$\sigma$. The grey solid line and black dashed line show the correlation function of \simba\ QSOs that provide the best match to the \protect\cite{White_2012} power law, whereby the QSOs are selected with the combined mass-luminosity criterion and with the luminosity cut, respectively (see \S~\ref{sec:sel_QSO} and \S~\ref{sec:haloes} for details). The plot demonstrates that these two selection methods are equivalent in the  $100 \hMpc$ \simba\ run.} \label{fig:corr_100}
\end{figure}

In this section, we provide a more detailed discussion on our selection method for QSO hosts. We begin by comparing the QSO sample selected with our fiducial technique based both on halo mass and luminosity of QSO hosts with the one obtained by applying a luminosity cut on BHs, without any reference to the mass of the host halo (see \S~\ref{sec:sel_QSO}).

Figure~\ref{fig:corr_100} shows the family of autocorrelation functions of central galaxies within the \simba\ $100 \hMpc$ run, obtained by varying the minimum luminosity $L_{\rm min}$  of the respective central BHs. The colour coding of the circles in Figure~\ref{fig:corr_100} allows identifying the autocorrelation function that corresponds to a specific value of $L_{\rm min}$. The black dotted line is the best-fit power-law to the observations of QSO clustering by \cite{White_2012}, and the shaded grey area around it represents the error around such power law within $1\sigma$. We now determine the optimal luminosity threshold by seeking the value of the BH accretion rate that corresponds to a luminosity $L_{\rm min}$ such that the autocorrelation function of galaxies hosting a BH with luminosity larger than $L_{\rm min}$ minimises the reduced $\chi^2$ when compared with the \cite{White_2012} best-fit power law. Such optimal correlation function is plotted with a black dashed line in Figure \ref{fig:corr_100}. As a reference, the grey solid line shows the optimal autocorrelation function obtained by our fiducial mass-and-luminosity selection criterion explained in \S~\ref{sec:sel_QSO}. We can clearly see that it coincides with the dashed black line, therefore the luminosity-only and luminosity-and-mass selection criteria explained in this section result in the selection of exactly the same sample of QSOs in the \simba\ $100 \hMpc$ run.

\begin{table}
\begin{center}
\begin{tabular}{lccc}
\hline
Simulation & \multicolumn{2}{c}{Fiducial} & Simplified \\
 & $\log\left(\frac{M_{\rm min}}{\rm M_{\odot}} \right)$ &  $\log\left(\frac{L_{\rm min}}{\rm erg \, s^{-1}} \right)$ & $\log\left(\frac{L_{\rm min}}{\rm erg \, s^{-1}} \right)$  \\
\hline
\simba\ $100 \hMpc$ & 12.7 & 45.3 & 45.3 \\
\simba\ $50 \hMpc$ & 12.8 & 45.4 & 44.2 \\
SFB + AGN Winds + Jets & 12.6 & 45.4 & 45.0 \\
SFB + AGN Winds & 12.6 & 45.4 & 45.1 \\
Stellar Feedback & 12.6 & 45.5 & 44.9 \\
No Feedback & 12.6 & 46.6 & 45.9\\
\hline
\end{tabular}
\caption{Optimal luminosity thresholds obtained with the fiducial method and the simplified luminosity-only selection criterion.}
\label{tab:selection}
\end{center}
\end{table}

If we repeat the same experiment for the $50 \hMpc$ runs, we find different optimal luminosity thresholds. For every run listed in the first column of Table~\ref{tab:selection}, we list the luminosity threshold (third column) corresponding to the optimal mass cut (second column) obtained with our fiducial selection criterion. In the fourth column we report the optimal luminosity floors given by the simpler luminosity-only technique. We notice that the differences among the luminosity thresholds\footnote{We remind the reader that we actually impose a threshold for the accretion rate of the central BH (see \S~\ref{sec:sel_QSO}). For runs that contain some form of AGN feedback, we can interpret it as a luminosity threshold because we can associate an AGN luminosity to the BH accretion rate by virtue of equation~\eqref{eq:L}. In runs without any AGN feedback prescription, the physical meaning of ``luminosity threshold'' is less straightforward. However, we can still associate a pseudo-luminosity $L$ to the accretion rate of a BH, which represents the luminosity of the hypothetical AGN powered by the BH in question if it were drawn from an analogous run with some form of AGN feedback. Though, we stress that our selection criteria can still be applied on all runs, with or without AGN feedback, because all of them include BH particles, hence the BH accretion rate is always well defined.} obtained with the fiducial criterion for the various $50 \hMpc$ runs stay within $0.2 \, \rm dex$, except for the no-feedback run. On the other hand, the simplified luminosity-only criterion exhibits a larger spread (up to $0.9 \, \rm dex$) in $L_{\rm min}$ across the $50 \hMpc$ runs endowed with at least stellar feedback. Moreover, in the \simba\ $50 \hMpc$ run the value of $L_{\rm min}$ is about one order of magnitude smaller than in its $100 \hMpc$ counterpart. On top of the smaller spread in $L_{\rm min}$ for most runs, the fiducial mass-and-luminosity criterion provides a better reduced $\chi^2$ when compared with \cite{White_2012} observations.

We thus conclude that the fiducial method is more robust, while the simplified selection criterion based solely on a luminosity cut tends to underestimate the optimal $L_{\rm min}$. The fact that for the \simba\ $100 \hMpc$ run the two methods give the same result suggests that the two techniques tend to agree as the volume of the simulation, and hence the statistics of available haloes, increases. Another advantage of the fiducial technique is that it enables a straightforward comparison with the results of other works in the literature where QSOs are selected in simulations via a halo mass cut only. 

Considering that for the $50 \hMpc$ runs the fiducial method provides us with mass thresholds differing by only 0.1 dex from the one obtained with the \simba\ $100 \hMpc$, we decided to impose the value of $10^{12.7} \, \rm M_{\odot}$ as the mass cut defining the luminosity threshold in all $50 \hMpc$ runs. We show in \S~\ref{sec:threshold} that such small differences have negligible impact on the final results of this work. 

\subsection{Satellite galaxies}
\label{sec:central}

In this work, only central galaxies can act as QSO hosts following our selection criteria (see \S~\ref{sec:sel_QSO}). This is motivated by the fact that halo model fits \citep{Conroy_2013} to observations of QSO clustering \citep{White_2008} indicated that the satellite fraction should be very low at the redshift of interest for our work. Furthermore, \cite{Richardson_2012} inferred a satellite fraction of $(7.4 \pm 1.4) \times 10^{-4}$ from observations of QSO clustering in the redshift range $0.4\lesssim z \lesssim 2.5$, and \cite{Kayo_2012} deduced a $0.054^{0.017}_{-0.016}$ satellite fraction from measurements of the small-scale clustering of QSOs in the range $0.6\lesssim z \lesssim 2.2$. On the other hand, \cite{Alam_2020} found a higher satellite fraction ($0.2-0.4$) for QSOs from the eBOSS survey, although at lower redshift ($0.7\leq z\leq 1.1$) with respect to our range of interest.

As a consistency check for our assumption, we explicitly verified that allowing satellite galaxies to act as QSO hosts in the \simba\ $50 \hMpc$ run would enlarge the resulting QSO sample by only 2 units ($7-8\%$). The resulting mean \lya flux contrast profile differs by less than $0.005$ over the full range of transverse distances probed. This is negligibly small compared to the error bars of \cite{Prochaska_2013} observations, and to other possible sources of uncertainty (see, e.g., \S~\ref{sec:threshold}, \S~\ref{sec:z-distribution}). Therefore, our approximation is well justified.

\section{Assessment of systematics in the analysis}
\label{sec:systematics}

To predict the mean \lya flux contrast around QSOs with the \simba\ suite of simulations, we inevitably had to make certain approximations and assumptions, which may in principle affect our results. In the next subsections we will examine the different possible sources of systematic errors, and quantify to what extent they affect the main conclusions of our work.

\subsection{Luminosity threshold}
\label{sec:threshold}
\begin{figure}
	\includegraphics[width=\columnwidth]{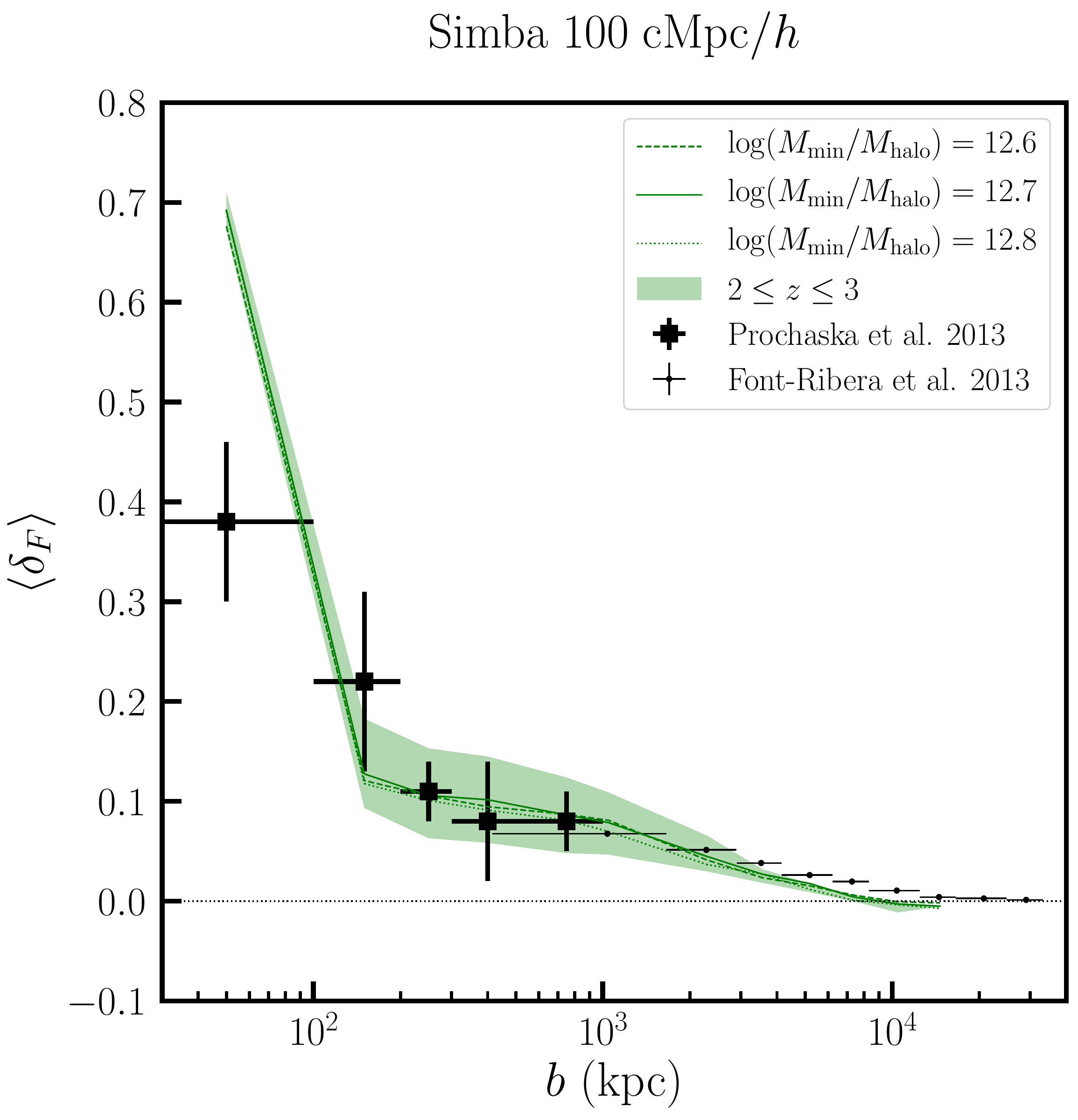}
    \caption{Mean \lya flux contrast profile around $z=2.4$ QSOs selected with initial mass cuts of $10^{12.6} \, \rm M_{\odot}$, $10^{12.7} \, \rm M_{\odot}$ (fiducial value), and $10^{12.8} \, \rm M_{\odot}$, corresponding to QSOs brighter than $10^{45.1} \, \rm erg/s$, $10^{45.3} \, \rm erg/s$, and $10^{45.5} \, \rm erg/s$, respectively. They are represented with the dotted, solid and dashed green lines, respectively. Differences of $\pm 0.1 \rm dex$ in the initial mass cut, translating into differences of $\pm 0.1 \rm dex$ in the QSO brightness, do not change the conclusions of this work. The green shaded area around the green solid line is delimited by the $\langle \delta_F \rangle$ profiles corresponding to a mass cut of $10^{12.7} \, \rm M_{\odot}$, and where all QSOs are at $z=2$ and $z=3$. Approximating the redshift distribution of QSOs with the median of the redshift range has a major impact on the resulting mean \lya flux contrast profile, however it does not affect the main conclusions of our work. }\label{fig:QSO-lum}
\end{figure}

As explained in \S~\ref{sec:sel_QSO} and \S~\ref{sec:haloes}, we select QSO hosts in \simba\ by choosing the haloes hosting the $N$ fastest accreting BHs, where $N$ is determined with mass-based selection arguments calibrated with independent observations. Although our methodology is more sophisticated than other methods generally adopted in the literature, we still need to assess the impact of the luminosity threshold on the resulting mean \lya flux contrast profile. 

In Figure \ref{fig:QSO-lum} we plot the $\langle \delta _{\rm F} \rangle$ profile obtained from the \simba\ $100 \hMpc$  simulation. The solid green line corresponds to the results given by our fiducial halo mass cut of $10^{12.7} \, \rm M_{\odot}$, which generates a sample of QSOs with luminosity $\gtrsim 10^{45.3} \, \rm erg \, s^{-1}$ (see \S~\ref{sec:haloes}). We change the mass cut by 0.1 dex, obtaining the dotted and dashed green lines for a mass floor of $10^{12.6} \, \rm M_{\odot}$ and $10^{12.8} \, \rm M_{\odot}$, respectively. The resulting QSO samples have luminosities above $10^{45.1} \, \rm erg/s$ and $10^{45.5} \, \rm erg/s$, respectively.

The differences among the various profiles amount to $\lesssim 0.01$ in the transverse distance range $100 \, \mathrm{kpc} \lesssim b \lesssim 1\Mpc$, whereas they are negligibly small ($<0.002$) on all other scales. We find differences of the same order of magnitude in the $50 \hMpc$ \simba\ runs as well. We conclude that errors of $\pm 0.1 \rm dex$ on the determination of the optimal mass cut (translating into $\sim \pm 0.2 \, \rm dex$ uncertainties in the resulting minimum luminosity of the QSO sample) would not change the conclusions of our work.

\subsection{Redshift distribution of QSOs}
\label{sec:z-distribution}

Throughout our analysis, we compute the mean \lya absorption profiles around QSOs at $z=2.4$, which is the median redshift of the foreground QSOs in the observations considered in this work. This is obviously a convenient simplifying approximation, given that the foreground QSOs observed by \cite{Prochaska_2013} and \cite{Font-Ribera_2013} are actually spread along the redshift range $2 \lesssim z \lesssim 3$. In fact, one should in principle consider multiple snapshots of the simulation within such redshift interval, with the aim of reproducing the observed QSO redshift distribution as faithfully as possible, and only at that point compute the resulting mean \lya flux contrast profile. Whereas most precise, this approach is considerably more time consuming and may be somewhat overzealous. We thus opt for a more efficient strategy to assess how much neglecting the redshift distribution of foreground QSOs impacts the predicted mean \lya flux contrast profile. 

We repeat the analysis of this work also at redshift $z=2$ and $z=3$, which bracket the redshift range of interest. The resulting $\langle \delta _{\rm F} \rangle$ profiles thus correspond to a hypothetical QSO sample whereby all objects are at $z=2$ and $z=3$, respectively. The absorption profile of the real QSO distribution will then be comprised between these two extremal profiles. The locus of all possible mean \lya flux profiles that are compatible with the foreground QSO distributions of \cite{Prochaska_2013} and \cite{Font-Ribera_2013} observations, as predicted by the \simba\ $100 \hMpc$ run, is shown in Figure \ref{fig:QSO-lum} as a green shaded area around the profile obtained for $z=2.4$ (green solid line).

Neglecting the spread in redshift of foreground QSOs has the highest impact in the range $100 \, \mathrm{kpc} \lesssim b \lesssim 1 \Mpc$, whereby the maximum error on the prediction of the mean \lya flux contrast profile amounts to $0.02-0.04$ ($25-30\%$), which is comparable with the differences among the various \simba\ runs on the same scales. On the other hand, even if we improved our selection method of QSOs in the simulations to match the redshift distribution of foreground QSOs in the observations, we would still be unable to discriminate among the different \simba\ runs in the range $100 \, \mathrm{kpc} \lesssim b \lesssim 1 \Mpc$ given the errors on the data. The error bars in the observations are expected to get smaller with upcoming surveys, and at that point it may be necessary to model the spread in redshift of foreground QSOs more carefully.

For $b \lesssim100 \kpc$ the maximum error is $\lesssim 0.02$, thus much smaller than the differences between the no-feedback run with respect to all other \simba\ runs, and also smaller than the discrepancies between \simba\ and \nyx\ or \illustris. Thus, selecting all QSOs from the snapshot corresponding to the median redshift of the sample does not affect our considerations about the physical properties of gas and the absorption profile within the innermost bin.

To summarise our findings for the small-scale regime ($b\lesssim 1\, \rm Mpc$), we conclude that although the redshift distribution of QSOs is a major contributor to the spread on the predicted mean \lya flux contrast within $1 \, \rm Mpc$ from QSOs, our simplified modelling does not affect our main conclusions on the CGM and CGM/IGM interface of QSOs. We also notice that our findings are consistent with the assessment of systematics performed by \cite{Sorini_2018} on \illustris\ and \nyx\ simulations. Also \cite{Rahmati_2013}, in a related work based on the \textsc{Eagle} suite of hydrodynamic simulations, concluded that the redshift distribution of foreground objects is the most important source of systematic errors in the modelling.

For completeness, in Figure~\ref{fig:QSO-lum} we assess the uncertainty due to the redshift distribution of QSOs up to the largest scales probed by our simulations. In this regime, the uncertainty drops from $\sim 0.03$ at $b\sim 1 \, \rm Mpc$ down to $\sim 0.01$ at $b\sim 10 \, \rm Mpc$. Even if we improved our modelling, hence reducing further the uncertainty, we would still be unable to reproduce the observations by \cite{Font-Ribera_2013} because of the already discussed limitations due to the box size of our simulations (see \S~\ref{sec:simba_100}-\ref{sec:feedback}). Considering the high precision of BOSS data, if we had a large enough simulation then it would be worth applying a fine modelling of the redshift distribution of QSOs. We leave this for future work.

\subsection{Position of the QSO host}

\label{sec:position}
\begin{figure}
	\includegraphics[width=\columnwidth]{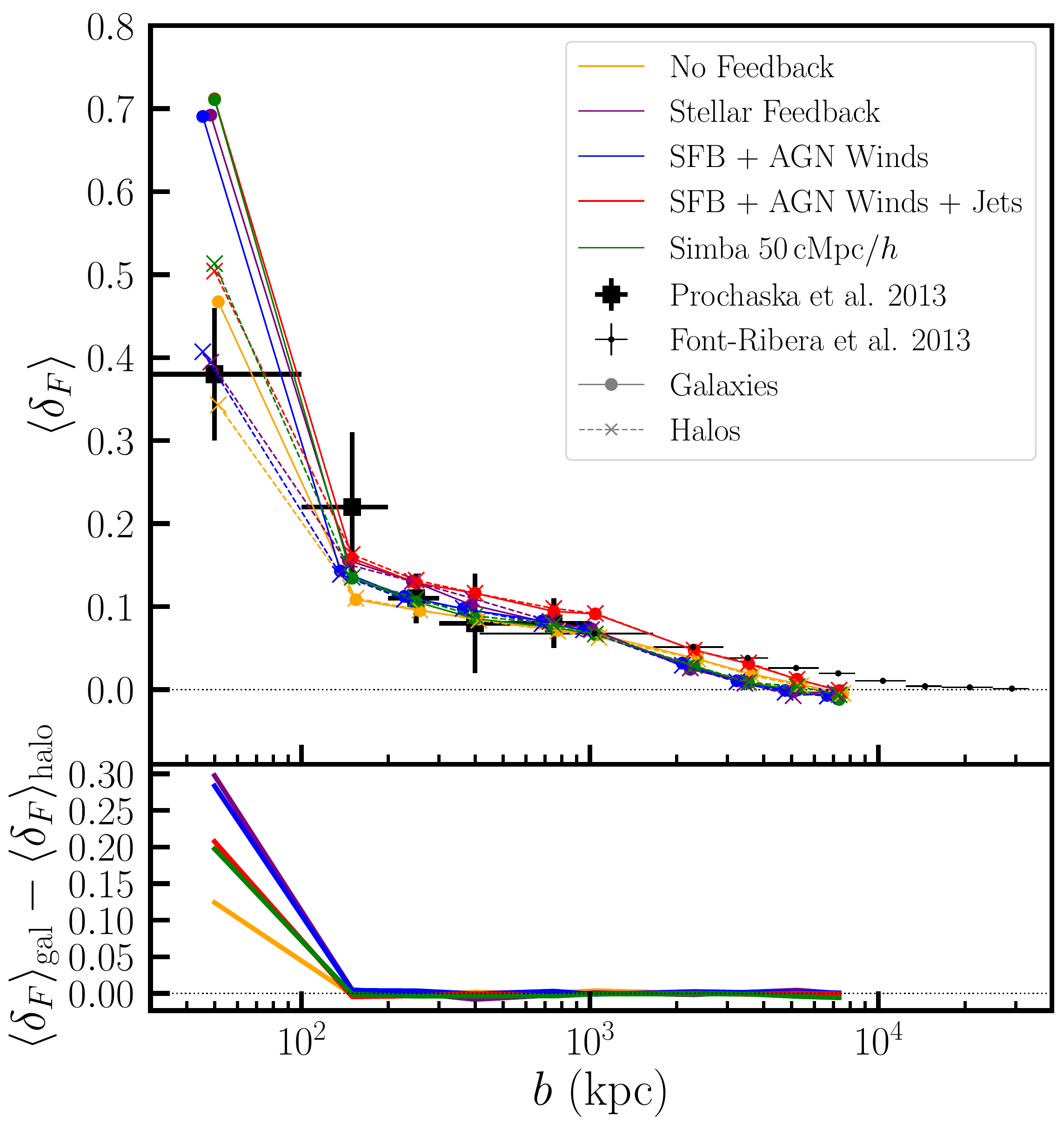}
    \caption{{\it Top panel:} Mean \lya flux contrast profile around QSOs taken from the $50 \hMpc$ \simba\ runs. Solid and dashed lines refer to profiles obtained by measuring the transverse distance of skewers from the centre of the host galaxy and host halo, respectively. The lines are colour-coded as in Figure \ref{fig:delta_F}. {\it Bottom panel}: Difference between the \lya flux contrast profiles where the transverse distance is measured from the host galaxy and the host halo, respectively. The colour coding of each line is the same as in Figure \ref{fig:delta_F}. The choice of the origin from which the transverse distance from the QSO is measured in the simulation has decisive impact on the mean \lya flux contrast in the innermost bin. }\label{fig:gal-halo}
\end{figure}

As we explained in \S~\ref{sec:skewers}, we draw skewers within different bins of transverse distance with respect to the QSOs  selected in \simba. Such distance is evaluated from the centre of the galaxies acting as QSO hosts in our work, and not from the centre of the parent haloes. This is possible because we cross-matched galaxies and haloes in post-processing with the \texttt{yt}-based package \textsc{Caesar}. On the contrary, our approach is obviously not applicable on simulations that do not include galaxy-formation physics. This is the reason why in previous work (such as \citealt{Sorini_2018}) the sample of skewers around QSOs had to be constructed by measuring transverse distances from the centre of the haloes hosting the QSOs, and not from the centre of the host galaxies.

In this section, we investigate whether the choice of the origin of the transverse distances of the skewers extracted from \simba\ affects the resulting mean \lya flux contrast profile. In the top panel of Figure \ref{fig:gal-halo} we plot the $\langle \delta _{\rm F} \rangle$ profiles obtained for all $50 \hMpc$ \simba\ runs when transverse distances of skewers are measured from the centre of mass of host galaxies (which is our fiducial choice), as computed by \textsc{Caesar}, using the same colour coding and marker styles as in Figure \ref{fig:delta_F}. Markers are connected with solid thin lines, to guide the eye. We also plot the analogous profiles obtained by evaluating transverse distances from the centre of mass of host haloes; such profiles follow the same colour coding and marker styles, but the points are connected with dashed lines. To highlight the impact of the choice of the origin of the transverse distance, in the bottom panel we plot the difference between the \lya flux contrast profiles where $b$ is measured from the centre of mass of the host galaxy and of the host halo, respectively. We adopt the same colour coding as in the top panel.

We notice that choosing the centre of the host galaxy rather than that of the host halo makes no difference for $b\gtrsim 100 \kpc$. On the contrary, for $b \lesssim 100 \kpc$, such a choice gives rise to differences up to $0.3$ in the mean \lya flux contrast. Indeed, we verified that the histogram of the distance between the centres of central QSO-hosting galaxies and of their parent haloes is peaked at $10-30 \kpc$ depending on the run of \simba\ considered, with $60-65\%$ of galaxy-parent halo pairs having $<50 \kpc$ distance\footnote{Offsets of this magnitude are not atypical in more massive haloes that formed more recently and are thus less relaxed, as shown by e.g. \cite{Sanderson_2009}, albeit at lower redshift.} in all runs. Such length scales are comparable with the size of the innermost bin of \cite{Prochaska_2013} observations. This is the reason why measuring transverse distances from the centre of the host galaxy rather than the host halo has a larger impact on $\langle \delta _{\rm F} \rangle$ near the QSO.

It is noteworthy that a careful definition of the origin of the transverse distances of skewers has a larger impact on the final results than other factors, such as the luminosity/mass threshold adopted for the selection of QSOs. Furthermore, the findings discussed in this section should be borne in mind when comparing results from different simulations, where other choices on the definition of the ``transverse distance from the QSO'' may have been made.

\subsection{Mean flux in the \lya forest}
\label{sec:mean_flux}

\begin{figure}
	\includegraphics[width=\columnwidth]{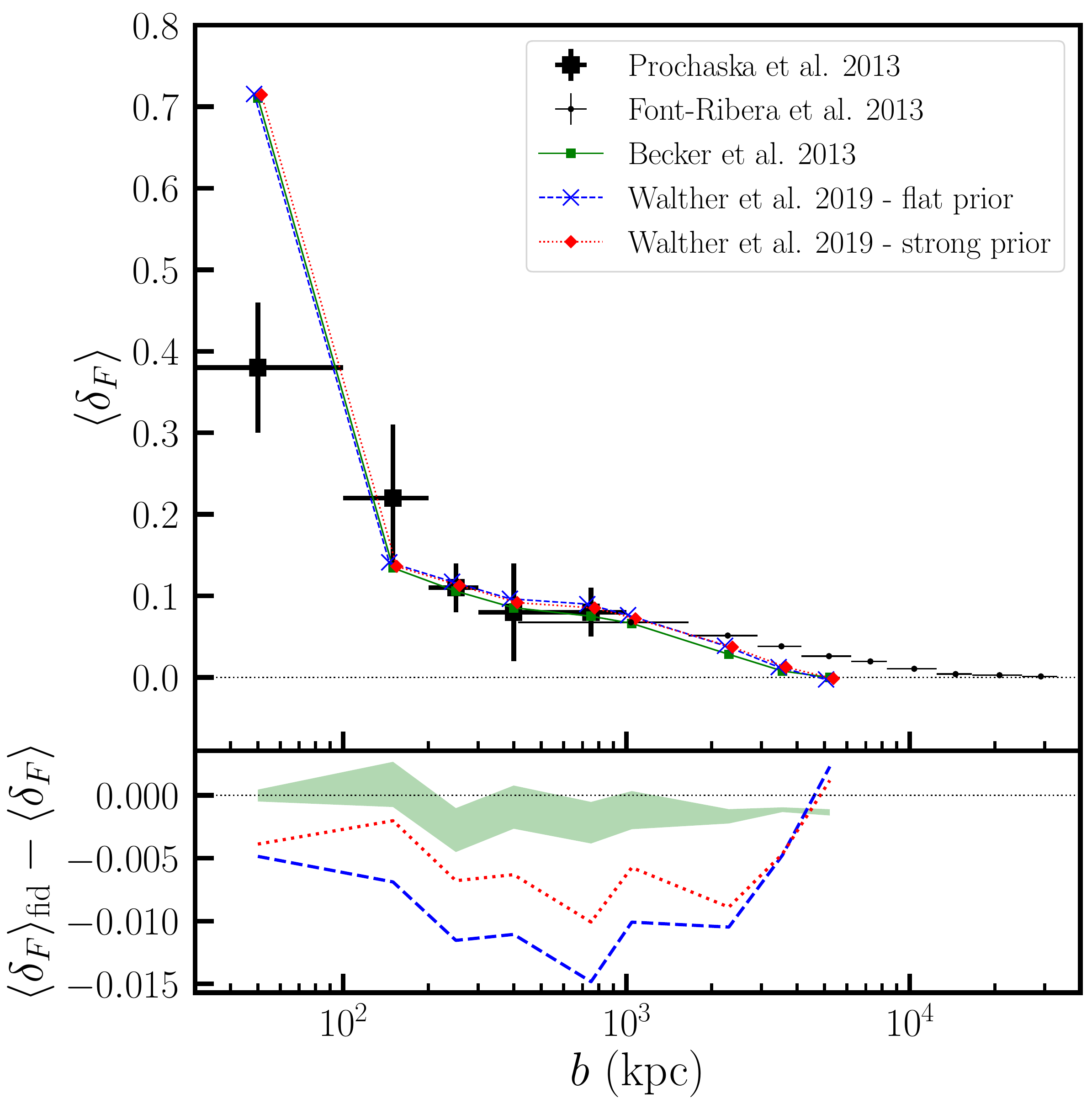}
    \caption{ {\it Top panel:} Mean \lya flux contrast profile around QSOs given by the \simba\ $50 \hMpc$ run after regulating the UVB to match the observations by \protect\citealt{Becker_2013_mean} (green squares connected with a green solid line), and \protect\citealt{Walther_2019} with flat and strong prior on the mean \lya flux (blue crosses connected with a blue dashed line and red diamonds linked by a red dotted line, respectively). 
    {\it Bottom panel:} Differences between the $\langle \delta_F \rangle$ profile obtained with our fiducial choice for the mean \lya flux in the IGM (i.e. \protect\citealt{Becker_2013_mean}) and by matching \protect\cite{Walther_2019}. The lines are colour coded as in the top panel. The green shaded area delimits the differences with respect to the fiducial $\langle \delta_F \rangle$ profile that would be obtained by choosing a value of the mean \lya flux in the IGM within $\pm 1 \sigma$ from \protect\cite{Becker_2013_mean} measurements. 
    The specific data set chosen to regulate the UVB does not significantly impact our results.}\label{fig:mean_flux}
\end{figure}

As explained in \S~\ref{sec:skewers}, before extracting \lya flux skewers around QSOs we regulate the UVB such that the mean \lya flux in the IGM at the median redshift of the observations matches the value measured by \cite{Becker_2013_mean}. We want to test how the error on these observations would propagate on our predictions of the mean \lya flux contrast.

In the top panel of Figure \ref{fig:mean_flux} we plot with green squares connected with a green solid line the \simba\ $100 \hMpc$ results obtained with the fiducial value of $0.8136$ for the mean \lya flux, inferred from \cite{Becker_2013_mean} observations at $z=2.4$. We then recompute our flux skewers after matching the UVB at $z=2.4$ to flux values within $1 \sigma$ ($0.0089$) from such value. The differences are always $\lesssim 0.003$, meaning that the errors on \cite{Becker_2013_mean} do not change the conclusions of this work.  

We also regulated the UVB to reproduce more recent measurements by \cite{Walther_2019}. The authors determine the mean \lya flux in the IGM by applying an MCMC on measurements of the power spectrum of the \lya forest \citep{Walther_2018}. The authors consider first a flat prior on the mean \lya flux, and then a ``strong'' Gaussian prior, obtaining $0.772^{+0.013}_{-0.012}$ and $0.799\pm0.008$ at $z=2.4$, respectively. We show the resulting mean \lya flux contrast profiles in Figure~\ref{fig:mean_flux} with blue crosses connected with a blue dashed line and with red diamonds linked by a red dotted line, respectively.

To facilitate the comparison between the different flux contrast profiles, in the bottom panel of Figure~\ref{fig:mean_flux} we plot the difference between the fiducial $\langle \delta_F \rangle$ profile (i.e., matched to \citealt{Becker_2013_mean}) and the profiles obtained by matching the \cite{Walther_2019} mean flux with flat and strong prior, adopting the same colour coding as in the top panel. The green shaded area delimits the differences expected in the $\langle \delta_F \rangle$ profile by choosing a value of the mean \lya flux in the IGM within $\pm 1 \sigma$ from \cite{Becker_2013_mean} measurements. The green shaded area is always consistent with zero within the LOS-to-LOS variance of the $\langle \delta_F \rangle$ profile ($\lesssim 0.003$ for $b> 100 \kpc$).

To summarise, all profiles are fully consistent with that obtained adopting \cite{Becker_2013_mean} measurements of the mean \lya flux, and the differences among the various profiles are within $0.015$ across all scales. We thus conclude that the choice of the data sets to match the mean \lya flux to has a marginal impact on our results, and does not alter our conclusions.

\section{Convergence test}
\label{sec:convergence}

\begin{figure}
	\includegraphics[width=\columnwidth]{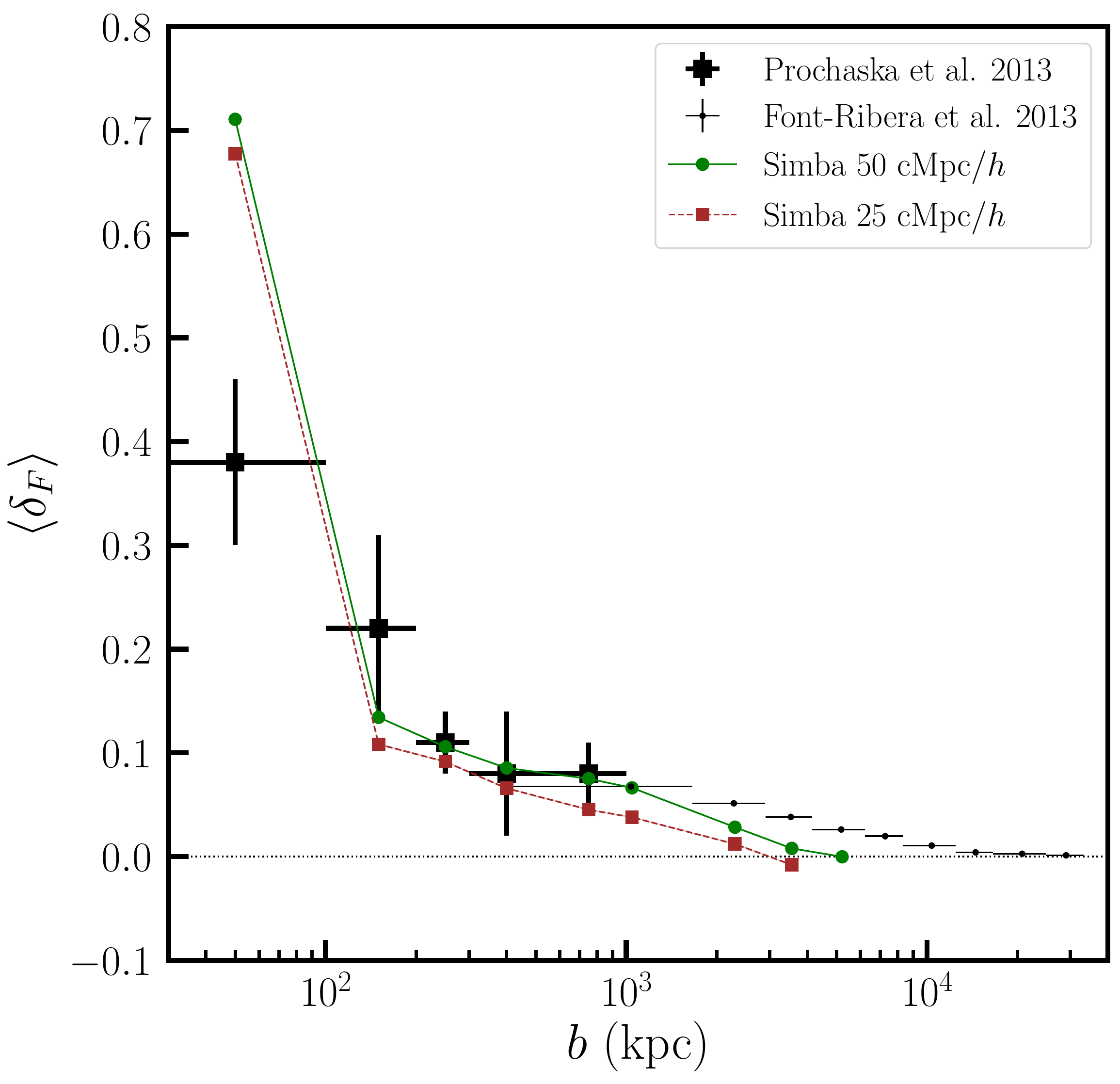}
    \caption{Mean \lya flux contrast profile around QSOs given by the \simba\ $50 \hMpc$ (green circles connected by the green solid line) and \simba\ $25 \hMpc$ (brown squares connected by the brown dashed line) runs. The latter run has twice the resolution of the former. Within $100 \kpc$, the two runs agree within $4.6\%$, and we can consider our results converged resolution wise.}\label{fig:convergence}
\end{figure}

By comparing the different \simba\ runs among themselves and with \nyx\ and \illustris, we showed that the most constraining transverse distance bin is $b<100 \kpc$. We already showed in \S~\ref{sec:feedback} that the \simba\ $50 \hMpc$ and \simba\ $100 \hMpc$ give the same predictions in this bin (see \S~\ref{sec:results}). Given that the resolution is critical at closer transverse separations, we now want to make sure that our results are converged also resolution-wise within the CGM of QSOs, and particularly in the aforementioned bin.

We computed the mean \lya flux contrast profile with the \simba\ $25 \hMpc$ run, which has a resolution eight times higher than the \simba\ $50 \hMpc$ run. In Figure~\ref{fig:convergence} we plot the results of the \simba\ $50 \hMpc$ and  \simba\ $25 \hMpc$ runs with green circles connected with a green solid line and brown squares linked by a brown dashed line, respectively. The differences between the two runs stay within $0.033$ across the whole range of scales, corresponding to a $4.6\%$ difference in the innermost bin. Given the magnitude of such differences, we can consider our results to be converged resolution wise. 

We caution that this conclusion is limited to scales $b\lesssim 700 \kpc$, comparable with the CGM size. Predictions of \simba\ $25\hMpc$ on scales $b\gtrsim 700 \kpc$ are probably not very reliable, as they are affected by the already discussed box-size effect (see \S~\ref{sec:simba_50}). Indeed, simulations with different box sizes cannot converge in $\langle \delta_F \rangle$ on the largest scales.

\bsp	
\label{lastpage}
\end{document}